\documentclass[journal,12pt,onecolumn,draftclsnofoot,]{IEEEtran}
\ifCLASSINFOpdf
\else
\fi

\usepackage{doi}
\usepackage{tikz}
\usepackage{graphicx}
\usepackage{mathtools}
\usepackage{romannum}
\usepackage{amsmath,amsfonts,amsthm,amssymb,mathrsfs}
\usepackage{type1cm}
\usepackage{lineno}
\usepackage{eso-pic}
\usepackage{color}
\usepackage{subfigure}
\usepackage{color}
\usepackage{fancyhdr}
\usepackage{lastpage}
\usepackage{adjustbox}
\usepackage{xfrac}
\usepackage{titling}
\usepackage[utf8]{inputenc}
\usepackage{float}
\usepackage{cite}

\DeclareGraphicsExtensions{.pdf,.jpeg,.png}
\graphicspath{{./img/}}

\makeatletter
\def\thanks#1{\protected@xdef\@thanks{\@thanks
        \protect\footnotetext{#1}}}
\makeatother

\begin{document}
\sloppy
%
\title{DOA Estimation in Nonuniform Sensor Noise}
%
%
%

\author{Majdoddin~Esfandiari,
        and~Sergiy~A.~Vorobyov,~\IEEEmembership{Fellow,~IEEE}
\thanks{M. Esfandiari and S. A. Vorobyov are with Dept. Signal Processing and Acoustics, Aalto University, PO Box 15400, 00076 Aalto, Finland. Emails: {majdoddin.esfandiari@aalto.fi; sergiy.vorobyov@aalto.fi}.
	Corresponding author is S. A. Vorobyov.} 
	\thanks{This work was supported in part by Academy of Finland under Research
Grant 319822.}}

%
%

\markboth{}%
{Esfandiari \MakeLowercase{\textit{et al.}}} 
%



\date{}
      \maketitle   
       \thispagestyle{empty}
        \makeatletter
         \let\ps@oldempty\ps@empty 
         \renewcommand\ps@empty \ps@headings
         \makeatother
        \rhead{\thepage}
\begin{abstract}
The problem of direction-of-arrival (DOA) estimation in the presence of nonuniform sensor noise is considered and a novel algorithm is developed. The algorithm consists of three phases. First, the diagonal nonuniform sensor noise covariance matrix is estimated using an iterative procedure that requires only few iterations to obtain an accurate estimate. The asymptotic variance of one iteration is derived for the proposed noise covariance estimator. Second, a forward-only rooting-based DOA estimator as well as its forward-backward averaging extension are developed for DOA estimation. The DOA estimators take advantage of using second-order statistics of signal subspace perturbation in constructing a weight matrix of a properly designed generalized least squares minimization problem. Despite the fact that these DOA estimators are iterative, only a few iterations are sufficient to reach accurate results. The asymptotic performance of these DOA estimators is also investigated. Third, a newly designed DOA selection strategy with reasonable computational cost is developed to select $L$ actual sources out of $2L$ candidates generated at the second phase. Numerical simulations are conducted in order to establish the considerable superiority of the proposed algorithm compared to the the existing state-of-the-art methods in challenging scenarios in both cases of uniform and nonuniform sensor noise.    
\end{abstract}


%
\IEEEpeerreviewmaketitle

\section{Introduction}
%
%
%
%
\IEEEPARstart{D}IRECTION-of-arrival (DOA) estimation is an active field of research for decades due to multiple traditional and new important applications and due to significance of source localization in many practical scenarios. The notable current applications of interest are, for example, wireless communication, automotive radar, and sonar where DOA estimation is an essential task \cite{van2004optimum, chung2014doa, krim1996two, upadhya2015array, upadhya2018low, hassanien2011transmit, khabbazibasmenj2014efficient, zhang2021gridless, ardah2021trice}. 

In the context of DOA estimation, several assumptions can be regarded concerning the structure of the second-order statistics of the observation noise. Most common assumptions are that the noise is uniform white and that the noise is nonuniform white. Moreover, a spatially block-correlated noise assumption may be more accurate in some applications \cite{Sergiy2005}. 

For the uniform white noise assumption, i.e., when the noise powers are identical across all array sensors, many well-known subspace-based methods such as MUSIC \cite{schmidt1986multiple, kim2012compressive, vallet2015performance}, root-MUSIC \cite{barabell1983improving, pesavento2000unitary, qian2013improved}, and ESPRIT \cite{roy1989esprit, haardt1995unitary} have been developed, and are based on decomposing the signal or its sample covariance matrix (SCM) into two disjoint subspaces of noise and signal. The popularity of the subspace-based methods over, for example, near optimal maximum likelihood (ML) estimators \cite{stoica1989music, stoica1990maximum, stoica1991performance}, comes from the fact that they achieve good estimation accuracy along with affordable computational complexity in contrast to the prohibitive computational cost of implementing ML estimators. In recent years, several works have proposed competitive algorithms to fill the performance gap between subspace-based methods and ML estimators including root-swap root-MUSIC \cite{shaghaghi2015subspace}, enhanced principal-singular-vector utilization for modal analysis (EPUMA) \cite{qian2016enhanced}, standard ESPRIT using generalized least squares (SE GLS) and unitary ESPRIT using generalized least squares (UE GLS) \cite{steinwandt2017generalized}, partial relaxation (PR)-based approaches \cite{trinh2018partial, trinh2020cramer}, root-clustering algorithm and root-certificate algorithm \cite{morency2018joint}. The aim of the aforementioned methods is to achieve an adequate performance in challenging scenarios like scarcity of available data samples, low signal-to-noise ratio (SNR), and/or presence of some correlated or even coherent sources. In \cite{shaghaghi2015subspace}, a new concept called \textit{root swap} has been introduced and recognized as the main reason behind the collapse of the root-MUSIC algorithm in finding the correct roots in the challenging environments. To remedy this phenomena, the root-swap root-MUSIC was proposed which identifies the correct roots associated with the sources' DOAs by exploiting the deterministic ML (DML) or stochastic ML (SML) \cite{stoica1990performance} objective functions instead of deciding based on the closeness of the absolute values of the roots to unity. In \cite{morency2018joint}, a new criterion for root selection based on the algebraic structure of the noise subspace has been proposed. Using this criterion and the relationship between the source localization problem and the problem of computing the approximate greatest common divisor (GCD) for polynomials, algorithms that learn the number of sources and estimate their locations have been proposed. In \cite{qian2016enhanced}, the authors have developed a method called EPUMA via solving a particular generalized least squares (GLS) problem by taking into account the second-order statistics of the estimated signal subspace, and also producing more DOA candidates than the number of sources and then selecting final DOAs using DML or SML cost functions. Similar to EPUMA, the authors of \cite{steinwandt2017generalized} have extended ESPRIT and unitary ESPRIT to their generalized versions by exploiting the signal subspace perturbation as the source of error in the shift invariance equation (SIE). The aforementioned error has been minimized using the GLS. Adopting the main idea of \cite{steinwandt2017generalized}, enhanced ESPRIT-based methods have been developed in \cite{esfandiari2021enhanced}, which first produces $2L$ DOA candidates for $L$ sources and then selects the final $L$ DOAs. The PR framework has been introduced in \cite{trinh2018partial} to relax the manifold structure of sources partially, resulting in four methods by applying the PR concept to four previously presented in the literature DOA estimation techniques. Note that most of the subspace based methods can be extended by means of using forward-backward averaging (FBA) \cite{linebarger1994efficient}, and forward-backward spatial smoothing (FBSS) \cite{pillai1989forward} techniques for the cases of correlated or coherent signals. 

The uniform noise assumption can be, however, violated in some practical scenarios, which appear increasingly more often, and the case of nonuniform white noise has drawn considerable attention in the last two decades. The deterministic ML estimator along with Cramer-Rao bound (CRB) for both deterministic and stochastic source models have been proposed in \cite{pesavento2001maximum}, while \cite{chen2008stochastic} have developed the stochastic ML algorithm for the case of nonuniform noise. A simple method has been devised in \cite{madurasinghe2005new} possessing less computational cost than ML estimators as well as improving the precision of the DOA estimates. In \cite{liao2016iterative}, the authors have developed two iterative methods referred to as iterative ML subspace estimation (IMLSE) and iterative least squares subspace estimation (ILSSE), which estimate the signal subspace and noise covariance matrices based on the ML and least squares (LS) criteria, respectively. Moreover, it has been shown in \cite{liao2016new} that the signal and noise subspaces are separable via applying the eigendecomposition (ED) of the so-called reduced covariance matrix for the case of uncorrelated sources, while a rank minimization approach has been suggested for the correlated sources case. The performance degradation caused by correlated and/or coherent sources can also be mitigated via using spatial smoothing \cite{wen2017spatial}, and covariance matrix differencing \cite{qi2007doa}. Moreover, bearing in mind the fact that the desired spatial directions can be modeled via sparse representation, sparse signal reconstruction (SSR) based methods have been developed \cite{stoica2010spice, he2014covariance, wang2018robust}. Recently, a method referred to as non-iterative subspace-based (NISB) \cite{esfandiari2019non} method has been proposed where the signal subspace and noise covariance matrix are identified by exploiting a two steps approach with the first step consisting of using the ED of the reduced covariance matrix \cite{liao2016new} for estimating noise covariance matrix judiciously, followed by applying the generalized eigendecomposition (GED) of the matrix pair of the SCM and estimated noise covariance matrix from the first step.

In this paper, a novel algorithm that consists of three phases is devised for addressing the problem of DOA estimation in the presence of nonuniform white noise. The proposed algorithm remarkably provides reliable estimates in scenarios with small snapshot size and/or relatively low SNR and/or closely located sources. In the first phase, noise covariance matrix is estimated by iterations via obtaining the noise subspace using the GED of two matrices, followed by updating noise covariance matrix using LS. Only few iterations are required to achieve an adequate estimate of the noise covariance matrix. Furthermore, the proposed noise covariance matrix estimator is applicable to sensor arrays with arbitrary geometry as well as any combination of correlated and uncorrelated sources. In addition, the asymptotic variance of the aforementioned estimator is derived. Using the noise covariance matrix estimate, in the second phase, the nonuniform noise DOA estimation problem is converted into the uniform noise DOA estimation problem by means of pre-whitening. Then a rooting-based approach that takes into account the signal subspace perturbation is used. It also utilizes the GLS for estimating the coefficients of the desired polynomial. To deal with possibly correlated sources, we also derive the FBA extension of the core DOA estimator. In addition, the asymptotic variance is derived for the DOA estimator under the high SNR assumption. In the last phase, a double number of DOA candidates is generated, and a DOA selection strategy is developed and used to pick the final DOAs. It is demonstrated by conducting extensive numerical simulations for various challenging setups that the proposed DOA estimation algorithm provides superior estimation accuracy both for the uniform and nonuniform noise cases compared to the existing state-of-the-art techniques.        

The rest of the paper is organized as follows. The signal model and problem formulation are given in Section~\ref{sec2}. In Section~\ref{sec3}, an iterative method for estimating diagonal nonuniform sensor noise covariance matrix is devised. The asymptotic variance of one iteration is also given for the proposed noise covariance estimator. The forward-only subspace-based DOA method as well as its FBA extension are developed for the case of nonuniform noise in Section~\ref{SubC}. Moreover, the asymptotic performance of the proposed DOA estimator is studied and a new DOA selection strategy is also designed to select final DOAs in the section. Numerical simulations are provided in Section~\ref{sec5}. The paper is concluded in Section~\ref{sec6}, and some technical derivations are given in Supplementary materials.

\textit{Notation:} Upper-case and lower-case bold-face letters denote matrices and vectors, respectively, while scalars are denoted by lower-case letters. The expectation, transpose, conjugate, and Hermitian transpose are denoted by $\mathbb{E} \{ \cdot \}$, $\{ \cdot \}^T$, $\{ \cdot \}^*$, and $\{ \cdot \}^H$, respectively, while $\| \cdot \|_2$, $\| \cdot \|_{\rm F}$, and $| \cdot |$ denote the Euclidean norm of a vector, the Frobenius norm of a matrix, and the absolute value of a scalar. If the argument is a set, $| \cdot |$ denotes the set cardinality. The Kronecker product is denoted by $\otimes$ and $\mathrm{trace} \{ \cdot \}$ stands for the trace of a square matrix. The $n \times n $ identity and exchange matrices are denoted by $\mathbf{I}_n$ and $\mathbf{J}_n$, respectively. The $n \times m$ matrix with all elements equal zero and $n \times 1$ zero vector are denoted as $\mathbf{0}_{m \times n}$ and $\mathbf{0}_{n}$, respectively. The $i$th entry of the vector $\boldsymbol{\pi}$ is denoted by $[\boldsymbol{\pi}]_{i}$. The $i$th row and $i$th column of the matrix $\boldsymbol{\Pi}$ are denoted by $[\boldsymbol{\Pi}]_{i,:}$ and $[\boldsymbol{\Pi}]_{:,i}$, respectively, while the entry in the interaction of the $i$th row and $j$th column is denoted as $[\boldsymbol{\Pi}]_{i,j}$. The operator $\mathrm{vec} \{ \cdot \}$ stacks the columns of a matrix into a long vector. The operator $\mathrm{diag} \{ \boldsymbol{\pi}\}$ generates a diagonal matrix by plugging the entries of the vector $ \boldsymbol{\pi}$ into its main diagonal, while the operator $\mathcal{D} \{ \boldsymbol{\Pi}\}$ creates a diagonal matrix by preserving the main diagonal of the matrix $\boldsymbol{\Pi}$ and setting all other entries to zero. The operator $\mathrm{DFT} \{ \boldsymbol{\pi}\}$ stands for the discrete Fourier transform (DFT) of the vector $\boldsymbol{\pi}$, while $\mathfrak{R} \{ \cdot \}$ returns the real part of the bracketed argument.

\section{SIGNAL MODEL}
\label{sec2}
Consider a uniform linear array (ULA) composed of $M$ omni-directional sensors receiving $L$ ($L < M$) independent narrowband signals radiated by $L$ sources. It is assumed that the sources are located in the far-field and have distinguished directions, denoted as $\theta_l$, $ l=1, \cdots, L$. Then, the signal observed by the sensor array at the time instant $t$ is written as      
\begin{flalign}
\mathbf{x}(t) = \mathbf{A}(\boldsymbol{\theta}) \mathbf{s}(t) + \mathbf{n}(t)
\label{eq1} \
\end{flalign}
where $s(t) \triangleq [s_1(t) \cdots s_L(t)]^T \in \mathbb{C}^L$ denotes the source signals, $\mathbf{n(t)} \in \mathbb{C}^M$ is the sensor noise vector, the source DOAs are stacked in the vector $\boldsymbol{\theta} \triangleq [\theta_1 \cdots \theta_L]^T$, $\mathbf{A}(\boldsymbol{\theta}) \triangleq [\mathbf{a}(\theta_1) \cdots \mathbf{a}(\theta_L)] $ denotes the array manifold whose $l$th column is the steering vector $\mathbf{a}(\theta_l) = [1 \ e^{-j2\pi \mathrm{sin}(\theta_l) d / \lambda} \ \cdots \ e^{-j2\pi (M-1) \mathrm{sin}(\theta_l) d / \lambda}]^T \in \mathbb{C}^{M}$ associated with $l$th DOA. Here $\lambda$ denotes the carrier wavelength and $d = \lambda/2$. For notation simplicity, $\mathbf{A}$ is used instead of $\mathbf{A}(\boldsymbol{\theta})$ hereafter, unless the argument of $\mathbf{A}$ is different from $\boldsymbol{\theta}$. 

The array covariance matrix can be written as
\begin{flalign}
\mathbf{R} \triangleq E\{\mathbf{x}(t) \mathbf{x}^H(t)\} = \mathbf{A} \mathbf{P} \mathbf{A}^H  + \mathbf{Q} 
\label{eq2} \
\end{flalign}
where $\mathbf{P} \in \mathbb{C}^{L \times L} $ and $ \mathbf{Q} \in \mathbb{R}^{M \times M}$, respectively,  stand for the signal and noise covariance matrices defined as 
\begin{flalign}
\mathbf{P} \triangleq E\{\mathbf{s}(t) \mathbf{s}^H(t)\}, \ \  \mathbf{Q} \triangleq E\{\mathbf{n}(t) \mathbf{n}^H(t)\}.
\label{eq300} \
\end{flalign}
Considering the case of nonuniform sensor noise that is spatially and temporally uncorrelated and zero-mean Gaussian, the noise covariance matrix is the following diagonal matrix
\begin{flalign}
\mathbf{Q} = \mathrm{diag} \{ [\sigma_1^2, \cdots, \sigma_M^2] \}
\label{eq301} \
\end{flalign}
where $\sigma_m^2$, $m=1, \cdots, M $ are the noise variances, which are not necessarily identical, i.e, $ \sigma_i^2 \neq \sigma_j^2$ for $ i \neq j$. When $\sigma_1^2 = \sigma_2^2 = \cdots = \sigma_M^2 = \sigma^2$, the noise covariance matrix is just a scaled identity matrix $ \mathbf{Q} = \sigma^2 \mathbf{I}_M$, i.e., the sensor noise is uniform. The latter case has been investigated to a great extent in the literature, while the former has been drawn more attention in recent years. The critical part of most of the methods designed for the nonuniform noise case is to estimate $ \mathbf{Q}$ efficiently, and the source DOA estimates are dependent on how precise the estimate of $\mathbf{Q}$ is. Moreover, when the number of available snapshots is small and/or some of the DOAs are closely located to each other and/or SNR is relatively low, the impact of estimated $\mathbf{Q}$ is more notable.    

Because $\mathbf{R}$ is unknown in practice, the SCM is considered, and it is given by
\begin{flalign}
\hat{\mathbf{R}} \triangleq \frac{1}{N} \displaystyle\sum_{t=1}^{N} \mathbf{x}(t) \mathbf{x}^H(t) = \frac{1}{N} \mathbf{X} \mathbf{X}^H .\  
\label{eq305} \
\end{flalign}
Here, the matrix signal notation is also used
\begin{flalign}
\mathbf{X} = \mathbf{A} \mathbf{S} + \mathbf{N}\  
\label{eq306} \
\end{flalign}
with $ \mathbf{X} \triangleq [\mathbf{x}(1) \cdots \mathbf{x}(N)]$, $ \mathbf{S} \triangleq~[\mathbf{s}(1) \cdots \mathbf{s}(N) ] $, $ \mathbf{N} \triangleq [\mathbf{n}(1) \cdots \mathbf{n}(N)]$, and $N$ being the number of snapshots. 

\section{NOISE COVARIANCE MATRIX ESTIMATION}
\label{sec3}
It is desirable that a method estimating $\mathbf{Q}$ would not require any knowledge of the true DOAs, while providing an acceptable accuracy with an affordable computational cost. Moreover, such method should be robust in extreme scenarios. Some examples of extreme scenarios are small sample size and presence of closely located sources. To devise such a method, we begin by multiplying both sides of \eqref{eq2} by $\mathbf{U} \in \mathbb{C}^{M \times (M-L)}$ which satisfies the following condition
\begin{flalign}
\mathbf{A}^H \mathbf{U} = \boldsymbol{0}_{L \times (M-L)}  
\label{eq307} \
\end{flalign}
where the constraint $\mathbf{U}^H \mathbf{U} = \mathbf{I}_{M-L}$ is also imposed for avoiding ambiguities in defining $\mathbf{U}$. It is clear then that the columns of $\mathbf{U}$ constitute an orthonormal basis for the noise subspace. As $\mathbf{A}$ is unknown, finding $\mathbf{U}$ for the case of nonuniform noise is not as simple as for the uniform noise case, when $\mathbf{U}$ is obtained by just calculating the eigenvectors of $\hat{\mathbf{R}}$. Multiplying \eqref{eq2} by $\mathbf{U}$, and exploiting \eqref{eq307}, it can be written that \cite{esfandiari2019non}
\begin{flalign}
\hat{\mathbf{R}} \mathbf{U} =  \mathbf{Q} \mathbf{U} 
\label{eq308} \
\end{flalign}
where $\mathbf{R}$ is replaced by the consistent estimate $\hat{\mathbf{R}}$. Then, it can be observed from \eqref{eq308} that for a given estimate of $\mathbf{Q}$, the columns of the best estimate of $\mathbf{U}$, denoted as $\hat{\mathbf{U}}$, are given by the $M-L$ eigenvectors associated with the $M-L$ smallest eigenvalues obtained after applying the GED to the pair of matrices $\hat{\mathbf{R}}$ and $\mathbf{Q}$. 

Furthermore, according to \eqref{eq308}, the LS approach can be employed to estimate $\mathbf{Q}$. In doing so, we formulate the following LS minimization problem with respect to $\mathbf{Q}$
\begin{align}
\hat{\mathbf{Q}} = \mathrm{arg}\displaystyle \min_{\mathbf{Q}} \|  (\hat{\mathbf{R}}-\mathbf{Q} ) \hat{\mathbf{U}} \|^2_{\rm F} \label{eq309}
\end{align}
where $\mathbf{U} $ is replaced by $\hat{\mathbf{U}}$. Problem \eqref{eq309} needs to be solved subject to the constraint on $\mathbf{Q}$, that is, $\mathbf{Q}$ is a diagonal matrix. The cost function of \eqref{eq309} can be expressed as
\begin{align}
f(\mathbf{Q}) &\triangleq  \left\|  (\hat{\mathbf{R}}-\mathbf{Q} ) \hat{\mathbf{U}} \right\|^2_{\rm F} 
= \mathrm{trace} \left\{ \left( (\hat{\mathbf{R}}-\mathbf{Q} ) \hat{\mathbf{U}} \right)  \left( (\hat{\mathbf{R}}-\mathbf{Q} ) \hat{\mathbf{U}} \right)^H \right\} \nonumber \\
&= \mathrm{trace} \left\{ \hat{\mathbf{U}} \hat{\mathbf{U}}^H \hat{\mathbf{R}}^2 \right\} - \mathrm{trace} \left\{ \hat{\mathbf{R}} \hat{\mathbf{U}} \hat{\mathbf{U}}^H \mathbf{Q} \right\} 
- \mathrm{trace} \left\{  \hat{\mathbf{U}} \hat{\mathbf{U}}^H \hat{\mathbf{R}} \mathbf{Q} \right\} \nonumber \\
&+ \mathrm{trace} \left\{  \hat{\mathbf{U}} \hat{\mathbf{U}}^H  \mathbf{Q}^2 \right\}
\label{eq310}
\end{align}
where the properties $ \| \mathbf{X} \|^2_{\rm F} = \mathrm{trace} \left\{ \mathbf{X} \mathbf{X}^H \right\}$, $\mathrm{trace} \left\{ \mathbf{X} \mathbf{Y} \right\} = \mathrm{trace} \left\{ \mathbf{Y} \mathbf{X} \right\} $, $\hat{\mathbf{R}} = \hat{\mathbf{R}}^H $, and $ \mathbf{Q} = \mathbf{Q}^H$ are employed. As elaborated in Appendix~\ref{APP1}, the partial derivative of \eqref{eq310} with respect to $ \mathbf{Q}$ is given as
\begin{flalign}
\frac{\partial f(\mathbf{Q})}{\partial \mathbf{Q}} = 2 \mathcal{D} \left\{ \hat{\mathbf{U}} \hat{\mathbf{U}}^H \right\} \mathbf{Q}  - \mathcal{D} \left\{ \hat{\mathbf{R}} \hat{\mathbf{U}} \hat{\mathbf{U}}^H + \hat{\mathbf{U}} \hat{\mathbf{U}}^H \hat{\mathbf{R}} \right\}. 
\label{eq311} 
\end{flalign}
Therefore, by equating \eqref{eq311} to zero, that is,
\begin{flalign}
 2 \mathcal{D} \left\{ \hat{\mathbf{U}} \hat{\mathbf{U}}^H \right\} \mathbf{Q}  - \mathcal{D} \left\{ \hat{\mathbf{R}} \hat{\mathbf{U}} \hat{\mathbf{U}}^H + \hat{\mathbf{U}} \hat{\mathbf{U}}^H \hat{\mathbf{R}} \right\} = 0 
\label{eq312} 
\end{flalign}
we find the optimal estimate of $ \mathbf{Q}$ as
\begin{flalign}
 \hat{\mathbf{Q}} = \frac{1}{2} \mathcal{D} \left\{ \hat{\mathbf{R}} \hat{\mathbf{U}} \hat{\mathbf{U}}^H + \hat{\mathbf{U}} \hat{\mathbf{U}}^H \hat{\mathbf{R}} \right\}  \mathcal{D} \left\{ \hat{\mathbf{U}} \hat{\mathbf{U}}^H \right\}^{-1} .
\label{eq313} 
\end{flalign}

According to \eqref{eq308} and \eqref{eq313}, which determine the dependencies between $\hat{\mathbf{Q}}$ and $\hat{\mathbf{U}}$, it is natural to select an iterative scheme for estimating $\hat{\mathbf{Q}}$ and $\hat{\mathbf{U}}$ in alternative manner. It starts with properly initializing $\hat{\mathbf{Q}}$, denoted as $ \hat{\mathbf{Q}}^{(0)}$, followed by estimating $\hat{\mathbf{U}}^{(0)}$ as the $M-L $ generalized eigenvectors of the pair $\{ \hat{\mathbf{R}}, \; \hat{\mathbf{Q}}^{(0)} \}$ associated with the $M-L$ smallest eigenvalues. Then, $ \hat{\mathbf{Q}}^{(1)}$ is obtained through \eqref{eq313} after replacing $\hat{\mathbf{U}}$ with $ \hat{\mathbf{U}}^{(0)}$. This procedure continues until a predefined stopping criterion is satisfied. Although any randomly generated diagonal matrix with positive diagonal entries can be used as $\hat{\mathbf{Q}}^{(0)}$, we suggest to initialize it as $\hat{\mathbf{Q}}^{(0)} = \mathcal{D} \{  \hat{\mathbf{R}}  \}$. 

\begin{table}[!h]
	\label{AL1}
	\begin{tabular}{l}
		\hline
		\textbf{Algorithm~1:} Noise Covariance Matrix Estimation
		\\ \hline
		1: Compute $\hat{\mathbf{R}} =  1 / N \sum_{t=1}^{N} \mathbf{x}(t) \mathbf{x}^H(t)$. \\
		2: Set $i=0$ and $\hat{\mathbf{Q}}^{(0)} = \mathcal{D} \{  \hat{\mathbf{R}} \}$. In addition, set the maximum number of iterations $i_{max} = 5$. \\
		\textbf{while} $i \leq i_{max}$ \\
		3: Carry out the GED of the pair of matrices $\{ \hat{\mathbf{R}}, \; \hat{\mathbf{Q}}^{(i)}) \}$ to obtain $\hat{\mathbf{U}}^{(i)}$  as the $M-L$ eigenvectors corresponding to the \\ $M-L$ smallest  eigenvalues. \\
		4: Calculate $\hat{\mathbf{Q}}^{(i+1)}$ using \eqref{eq313}. \\
		5: set $i=i+1$. \\
		\textbf{end} \\
		\hline
	\end{tabular}
\end{table}

The steps of the proposed method for noise covariance estimation are outlined in Algorithm~1. Additionally, to demonstrate the quality of the proposed noise covariance estimation method in \eqref{eq313}, its asymptotic mean square error (MSE) is found in the following proposition. 

\textit{Proposition 1:} The asymptotic variance of estimating each diagonal element of $\hat{\mathbf{Q}}$ in \eqref{eq313}, given a particular $\hat{\mathbf{U}}$, is
\begin{flalign}
\mathbb{E} \left\{ \left(\Delta \sigma_m^2 \right)^2 \right\} &= \left( \frac{ [\mathbf{R}]_{mm}}{2 N \tau_m^2} \right) \mathfrak{R} \left\{ \mathbf{v}_m^H \mathbf{R} \left( \mathbf{v}_m + N \mathbf{v}_m^* \right)\right\}
\label{eq414} 
\end{flalign}
where $\tau_m \triangleq [\hat{\mathbf{U}} \hat{\mathbf{U}}^H]_{mm}$ and $\mathbf{v}_m \triangleq [\hat{\mathbf{U}} \hat{\mathbf{U}}^H]_{:,m}$.

\textit{Proof:} See Appendix~\ref{APP2}. \  \ \ \  \ \ \ \ \ \  \ \ \  \ \ \  \ \ \ \ \ \  \ \ \  \ \ \    \ \ \ \ \ \  \ \ \  \ \ \  \ \ \ \ \ \  \ \  \  \ \ \  \ \ \ \ \ \  \ \ \  \ \ \  \ \ \ \ \ \  \ \ \ \ \ \ \ \ \ $\blacksquare$     

\textbf{Remark 1}: The estimate of $\hat{\mathbf{Q}}$ in \eqref{eq313} is an alternative representation of the power domain (PD) method \cite{madurasinghe2005new}, that is,
\begin{flalign}
\sigma_m^2 = \frac{\left(  \mathbf{d}_m^T \mathbf{P}_{\mathbf{A}} \hat{\mathbf{r}}_m + \hat{\mathbf{r}}_m^H \mathbf{P}_{\mathbf{A}} \mathbf{d}_m \right)}{2 \mathbf{d}_m^T \mathbf{P}_{\mathbf{A}} \mathbf{d}_m}, \ \ \  m=1, \cdots, M 
\label{eq314} 
\end{flalign}
where $\mathbf{d}_m \in \mathbb{R}^{M}$ denotes a vector with one on the $m$th position and zero elsewhere, $\hat{\mathbf{r}}_m$ is the $m$th column of $\hat{\mathbf{R}}$, and $\mathbf{P}_{\mathbf{A}} \in \mathbb{C}^{M \times M }$ is the orthogonal projection matrix of the signal subspace, i.e., $\mathbf{P}_{\mathbf{A}} = \mathbf{I}_M - \mathbf{A} (\mathbf{A}^H \mathbf{A})^{-1} \mathbf{A}^H$. The main difference between \eqref{eq313} and \eqref{eq314} is the use of $\hat{\mathbf{U}} \hat{\mathbf{U}}^H$ as a proper estimate of $\mathbf{P}$ instead of conducting a multidimensional search for finding an ML estimate as in \cite{madurasinghe2005new}. It is well-known that performing such multidimensional search is a very computationally demanding task. Therefore, the proposed method is advantageous compared to PD technique in the sense that it requires significantly less computations. 

\textbf{Remark 2}: As the actual $\mathbf{Q}$ is unknown, it is more accurate to modify \eqref{eq308} for the $i$th iteration as
\begin{flalign}
\hat{\mathbf{R}} \hat{\mathbf{U}}^{(i)} =  \hat{\mathbf{Q}}^{(i)} \hat{\mathbf{U}}^{(i)} \hat{\boldsymbol{\Lambda}}^{(i)} 
\label{eq315} \
\end{flalign}
where $\hat{\boldsymbol{\Lambda}}^{(i)} \in \mathbb{R}^{(M-L) \times 
(M-L)}$ is a diagonal matrix containing the generalized eigenvalues corresponding to the columns of $\hat{\mathbf{U}}^{(i)}$. As a consequence, the relationship between $\hat{\mathbf{Q}}^{(i+1)}$ and $\hat{\mathbf{Q}}^{(i)}$ can be obtained via exploiting \eqref{eq313} and \eqref{eq315} as follows
\begin{flalign}
 \hat{\mathbf{Q}}^{(i\!+\!1)} =& \frac{ \mathcal{D} \! \left\{ \hat{\mathbf{R}} \hat{\mathbf{U}}^{(i)} \hat{\mathbf{U}}^{(i)H} \!+\! \hat{\mathbf{U}}^{(i)} \hat{\mathbf{U}}^{(i)H} \hat{\mathbf{R}} \right\} \! \mathcal{D} \! \left\{ \hat{\mathbf{U}}^{(i)} \hat{\mathbf{U}}^{(i)H} \right\}^{-1} }{2}  \nonumber \\
 =& \mathcal{D} \left\{ \hat{\mathbf{U}}^{(i)} \hat{\boldsymbol{\Lambda}}^{(i)} \hat{\mathbf{U}}^{(i)H} \right\} \mathcal{D} \left\{ \hat{\mathbf{U}}^{(i)} \hat{\mathbf{U}}^{(i)H} \right\}^{-1} \hat{\mathbf{Q}}^{(i)} .
\label{eq316} 
\end{flalign}
Here the properties $\hat{\mathbf{Q}}^{(i)} = (\hat{\mathbf{Q}}^{(i)})^H$ and $\hat{\boldsymbol{\Lambda}}^{(i)} = (\hat{\boldsymbol{\Lambda}}^{(i)})^H$ have also been used. Hence, based on \eqref{eq316}, two consecutive estimates $\hat{\mathbf{Q}}^{(i)}$ and $\hat{\mathbf{Q}}^{(i+1)}$ are linked to each other through a diagonal weighting matrix $\mathbf{W}_Q^{(i)}$, that is, \begin{flalign}
\mathbf{W}_Q^{(i)} \triangleq \mathcal{D} \{ \hat{\mathbf{U}}^{(i)} \hat{\boldsymbol{\Lambda}}^{(i)} \hat{\mathbf{U}}^{(i)H} \} \mathcal{D} \{ \hat{\mathbf{U}}^{(i)} \hat{\mathbf{U}}^{(i)H} \}^{-1} .
\label{eq317} 
\end{flalign}   

\textbf{Remark 3}: The sufficient number of iterations for achieving a precise estimate of the noise covariance matrix for $\hat{\mathbf{Q}}^{(0)} = \mathcal{D} \{  \hat{\mathbf{R}}  \}$ is 3--5 as will be shown in simulation section.

\section{SUBSPACE-BASED DOA ESTIMATION VIA GLS}
\label{SubC}

\subsection{Forward-only Algorithm for DOA Estimation}
\label{sub31}
Using the estimate $\hat{\mathbf{Q}}$ obtained based on Algorithm~{1}, the received signal can be preprocessed by multiplying \eqref{eq306} by $\hat{\mathbf{Q}}^{-\frac{1}{2}}$ to enforce the uniform noise. Thus, the received signal becomes
\begin{flalign}
\overline{\mathbf{X}} \triangleq \hat{\mathbf{Q}}^{-\frac{1}{2}} \mathbf{X} = \hat{\mathbf{Q}}^{-\frac{1}{2}} \mathbf{A} \mathbf{S} + \hat{\mathbf{Q}}^{-\frac{1}{2}} \mathbf{N} = \hat{\mathbf{Q}}^{-\frac{1}{2}} \mathbf{A} \mathbf{S} + \overline{\mathbf{N}}  
\label{eq329} \
\end{flalign}
where the columns of $\overline{\mathbf{N}}$ are Gaussian random vectors with zero mean and the covariance matrix is $\mathbf{I}_M$.\footnote{In fact, a more accurate $\hat{\mathbf{Q}}$ leads to the covariance matrix of $\overline{\mathbf{N}}$ being closer to $\mathbf{I}_M$.} 

The truncated SVD of $\overline{\mathbf{X}}$ is given as
\begin{flalign}
\overline{\mathbf{X}} =  \mathbf{U}_{\rm s} \boldsymbol{\Sigma}_{\rm s} \mathbf{V}_{\rm s}^H
\label{eq330} 
\end{flalign}
where $\mathbf{U}_{\rm s} \in \mathbb{C}^{M \times L }$ and $ \mathbf{V}_{\rm s} \in \mathbb{C}^{N \times L }$ denote respectively the left and right singular vectors associated with the $L$ principal singular values on the diagonal of $\boldsymbol{\Sigma}_{\rm s} \in \mathbb{R}^{L \times L }$.

According to \eqref{eq329} and \eqref{eq330}, the columns of $\mathbf{U}_{\rm s}$, denoted as $ \mathbf{u}_{p}$ for $p = 1, \cdots, L$, and the columns of $\hat{\mathbf{Q}}^{-\frac{1}{2}} \mathbf{A}$ span the same column space, i.e., $\mathrm{span} (\mathbf{U}_{\rm s}) = \mathrm{span} (\hat{\mathbf{Q}}^{-\frac{1}{2}} \mathbf{A})$. In other words, $\mathbf{U}_{\rm s}$ and $\hat{\mathbf{Q}}^{-\frac{1}{2}} \mathbf{A}$ are related as
\begin{flalign}
 \mathbf{U}_{\rm s} = \hat{\mathbf{Q}}^{-\frac{1}{2}} \mathbf{A} \mathbf{G}
\label{eq331} 
\end{flalign}
where $\mathbf{G} \in \mathbb{C}^{L \times L }$ is a non-singular matrix. Consequently, multiplying \eqref{eq331} by $\hat{\mathbf{Q}}^{\frac{1}{2}}$, we have 
\begin{flalign}
 \widetilde{\mathbf{U}}_{\rm s} = \mathbf{A} \mathbf{G}     
\label{eq332} 
\end{flalign}
where $\widetilde{\mathbf{U}}_{\rm s} \triangleq \hat{\mathbf{Q}}^{\frac{1}{2}} \mathbf{U}_{\rm s} \in \mathbb{C}^{M \times L}$. 

Using \eqref{eq332}, the $m$th entry of the $p$th column of $\widetilde{\mathbf{U}}_{\rm s}$, denoted as $ [\widetilde{\mathbf{u}}_{p}]_m$, can be written as
\begin{flalign}
[\widetilde{\mathbf{u}}_p]_m  =& [\mathbf{A}]_{m,:} \ \mathbf{g}_p = \displaystyle\sum_{l=1}^{L} [\mathbf{g}_p]_l \ e^{-j2\pi d \mathrm{sin}(\theta_l) (m-1) /\lambda}, \nonumber \\
 &p=1, \cdots, L, \ m=1, \cdots, M \  
\label{eq333} \
\end{flalign}
where $\mathbf{g}_p \in \mathbb{C}^{L }$ is the $p$th column of $\mathbf{G}$. 

Next, the DFT can be applied to each column of $\widetilde{\mathbf{U}}_{\rm s}$ as, for example, in \cite{provencher2011parameters}. Using \eqref{eq333} and the definition of DFT, the $k$th bin of the DFT of $\widetilde{\mathbf{u}}_p$ can be expressed as  
\begin{flalign}
[\bar{\mathbf{u}}_p]_k = \displaystyle\sum_{l=1}^{L} [\mathbf{g}_p]_l & \ \frac{1-e^{jM\beta_l}}{1-e^{-j \frac{2 \pi k}{M}}e^{j\beta_l}} = \displaystyle\sum_{l=1}^{L} \frac{\alpha_l}{1-\gamma_l W_M^k} , \nonumber \\
& \ k=1, \cdots, M \  
\label{eq334} \
\end{flalign} 
where $\bar{\mathbf{u}}_p \triangleq \mathrm{DFT}\{\widetilde{\mathbf{u}}_p\} = \mathbf{W}_{D} \widetilde{\mathbf{u}}_p, \;  p=1, \cdots, L $ and
\begin{flalign}
\mathbf{W}_{D} &= \begin{bmatrix}
1 & 1 & \cdots & 1 \\
1 & e^{\frac{-j2 \pi}{M} } & \cdots & e^{\frac{-j2 \pi (M-1)}{M} } \\
\vdots  & \vdots  & \ddots & \vdots  \\
1 & e^{\frac{-j2 \pi (M-1)}{M} } & \cdots & e^{\frac{-j2 \pi (M-1)(M-1)}{M} }
\end{bmatrix} .
\nonumber 
\end{flalign}
In addition, $\alpha_l \triangleq [\mathbf{g}_p]_l \left(1-e^{j M\beta_l} \right) $, $\gamma_l \triangleq e^{j\beta_l}$, $ \beta_l \triangleq -2 \pi d \ \mathrm{sin}(\theta_l) / \lambda$ , and $W_M^k \triangleq e^{-j \frac{2 \pi k}{M}}$ in \eqref{eq334}. Unifying the $L$ rational functions into one, we recast \eqref{eq334} as
\begin{flalign}
[\bar{\mathbf{u}}_p]_k = \frac{\displaystyle\sum_{l=1}^{L} \alpha_l \displaystyle\prod_{\substack{v=1 \\ v \neq l }}^{L} (1-\gamma_v W_M^k)}{\displaystyle\prod_{l=1 }^{L} (1-\gamma_l W_M^k)} \  
\label{eq337} \
\end{flalign}
where the common denominator is the product of the $L$ denominators of each rational function. Noticing the special structure of \eqref{eq337}, the nominator and denominator can be expanded as two polynomials of degrees $L-1$ and $L$, respectively, that is, 
\begin{flalign}
\displaystyle\sum_{l=1}^{L} \alpha_l \displaystyle\prod_{\substack{v=1 \\ v \neq l }}^{L} (1-\gamma_v W_M^k) = \displaystyle\sum_{l=1}^{L} b_{pl} (W_M^k)^{l-1} = \overline{\mathbf{w}}_k^\intercal \mathbf{b}_p
\label{eq338} \\
\displaystyle\prod_{l=1 }^{L} (1-\gamma_l W_M^k) = 1+\displaystyle\sum_{l=1}^{L} a_l (W_M^k)^{l} = 1 + \mathbf{w}_k^\intercal \mathbf{a}    
\label{eq339} 
\end{flalign}
where $\overline{\mathbf{w}}_k \triangleq [1 \  W_M^k \ (W_M^k)^2 \ \cdots \ (W_M^k)^{L-1}]^T$, $\mathbf{w}_k \triangleq [W_M^k \ (W_M^k)^2 \ (W_M^k)^3 \ \cdots \ (W_M^k)^{L}]^T$, $\mathbf{b}_p \triangleq [b_{p1} \ \cdots \ b_{pL} ]^T$, and $\mathbf{a} \triangleq [a_1 \ \cdots \ a_L ]^T$.
Note that neither the entries of $\mathbf{b}_p$'s nor the entries of $\mathbf{a}$ are dependent to $k$. 

It can be seen from \eqref{eq339} that estimating $\mathbf{a}$ is the key for finding $\gamma_l$'s, $l=1, \cdots, L$, since $\gamma_l$'s, $l=1, \cdots, L$ are the roots of the polynomial defined by the entries of $\mathbf{a}$ as
\begin{flalign}
\gamma^L + \displaystyle\sum_{l=1}^{L} [\mathbf{a}]_l \ \gamma^{L-l} = 0.  \  
\label{eq343} \
\end{flalign}
Finally, knowing $\gamma_l$'s, $\theta_l$'s can be extracted using the relation $\theta_l = \mathrm{arcsin} (  -\frac{\beta_l \lambda}{2 \pi d} )$ where $\beta_l$ is the phase argument of $\gamma_l$ as it is defined above. 

Thus, the objective now is to find an estimate of $\mathbf{a}$, denoted as $\hat{\mathbf{a}}$. Multiplying both sides of \eqref{eq337} by the denominator, and using \eqref{eq338} and \eqref{eq339}, we have 
\begin{flalign}
[\bar{\mathbf{u}}_p]_k (1 + \mathbf{w}_k^\intercal \mathbf{a}) = \overline{\mathbf{w}}_k^\intercal \mathbf{b}_p . \  
\label{eq344} \
\end{flalign}
Piling up all the equations that can be generated for $k=1, \cdots, M$ based on \eqref{eq344}, we can write 
\begin{flalign}
\bar{\mathbf{u}}_p + \mathrm{diag}(\bar{\mathbf{u}}_p) \mathbf{W}_{a} \mathbf{a} = \overline{\mathbf{W}} \mathbf{b}_p, \ \ \ \ p=1, \cdots, L \  
\label{eq345} \
\end{flalign}
where $\mathbf{W}_{a} \triangleq [\mathbf{w}_1 \mathbf{w}_2 \cdots \mathbf{w}_M ]^T \in \mathbb{C}^{M \times L }$ and $\overline{\mathbf{W}} \triangleq  [\overline{\mathbf{w}}_1 \overline{\mathbf{w}}_2 \cdots \overline{\mathbf{w}}_M ]^T \in \mathbb{C}^{M \times L }$. 

We suggest to generate two sets of estimates for $\theta_l$'s as the most probable candidates first, and then pick up the best candidates via using a proper selection criteria. In doing so, we introduce the selection matrix $\mathbf{Z}_{\mathcal{I}} \in \mathbb{R}^{|\mathcal{I} | \times M }$. Here $\mathcal{I}$ denotes the set containing the indices of the selected equations. The matrix $\mathbf{Z}_{\mathcal{I}}$ is used to consider different sets of \eqref{eq345} for estimating $\mathbf{a}$. Since $\mathbf{Z}_{\mathcal{I}}$ is a selection matrix, all the entries of the $i$th row of $\mathbf{Z}_{\mathcal{I}}$ are zeros except one entry whose index is the $i$th member of $\mathcal{I}$. The only nonzero entry of each row is set to 1. The proposed method for selecting two sets of indices to serve as $\mathcal{I}$ will be clarified in the sequel. 

Given a particular set $\mathcal{I}$, the selected subset of \eqref{eq345} can be written as
\begin{flalign}
\mathbf{Z}_{\mathcal{I}} \bar{\mathbf{u}}_p + &\mathrm{diag}(\mathbf{Z}_{\mathcal{I}} \bar{\mathbf{u}}_p) \mathbf{Z}_{\mathcal{I}} \mathbf{W}_{a} \mathbf{a} = \mathbf{Z}_{\mathcal{I}} \overline{\mathbf{W}} \mathbf{b}_p, \quad p=1, \cdots, L.
\label{eq348} 
\end{flalign}
To avoid estimating $\mathbf{b}_p$'s in \eqref{eq348}, $\mathbf{B} \in \mathbb{C}^{|\mathcal{I}| \times (|\mathcal{I}|-L) }$ can be obtained using SVD such that $\mathbf{B}^H  \widetilde{{\mathbf{Z}}}_{\mathcal{I}} = \mathbf{0}_{(|\mathcal{I}|-L) \times L}$ with $ \widetilde{{\mathbf{Z}}}_{\mathcal{I}} \triangleq \mathbf{Z}_{\mathcal{I}} \overline{\mathbf{W}} \in \mathbb{C}^{|\mathcal{I}| \times L }$. Thus, multiplying both sides of \eqref{eq348} by $\mathbf{B}^H$ and using the property $\mathbf{B}^H  \widetilde{{\mathbf{Z}}}_{\mathcal{I}} = \mathbf{0}_{(|\mathcal{I}|-L) \times L}$, yields    
\begin{flalign}
\mathbf{B}^H (\mathbf{Z}_{\mathcal{I}} \bar{\mathbf{u}}_p + &\mathrm{diag}(\mathbf{Z}_{\mathcal{I}} \bar{\mathbf{u}}_p) \mathbf{Z}_{\mathcal{I}} \mathbf{W}_{a} \mathbf{a}) = \mathbf{0}_{(|\mathcal{I}|-L)}, \; p=1, \cdots, L.
\label{eq349} \
\end{flalign}
The impact of $\mathbf{b}_p$'s is eliminated in \eqref{eq349} independent of the value of $p$ because of the definition of $\widetilde{{\mathbf{Z}}}_{\mathcal{I}}$, which is independent of $p$. This property enables us to combine the $L$ sets of linear equations generated via \eqref{eq349}. Rearranging the terms in \eqref{eq349}, we get the following system of linear equations $\mathbf{H}_p \mathbf{a} = \mathbf{h}_p, \; p=1, \cdots, L$, where $\mathbf{H}_p \triangleq \mathbf{B}^H \mathrm{diag} (\mathbf{Z}_{\mathcal{I}} \bar{\mathbf{u}}_p) \mathbf{Z}_{\mathcal{I}} \mathbf{W}_{a} \in \mathbb{C}^{(|\mathcal{I}|-L) \times L }$ and $\mathbf{h}_p \triangleq -\mathbf{B}^H \mathbf{Z}_{\mathcal{I}} \bar{\mathbf{u}}_p \in \mathbb{C}^{(|\mathcal{I}|-L) }$.
Stacking the $L$ matrices $\mathbf{H}_p$ and the $L$ vectors $\mathbf{h}_p$ into a larger matrix $\mathbf{H}$ and a longer vector $\mathbf{h}$, respectively, we have 
\begin{flalign}
\mathbf{H} \mathbf{a} = \mathbf{h}\  
\label{eq353} \
\end{flalign}
where $\mathbf{H} \triangleq [\mathbf{H}_1^T \cdots \mathbf{H}_L^T]^T \in \mathbb{C}^{L(|\mathcal{I}|-L) \times L }$ and $\mathbf{h} \triangleq [\mathbf{h}_1^T \cdots \mathbf{h}_L^T]^T \in \mathbb{C}^{L(|\mathcal{I}|-L)}$.

However, only an estimate of $\mathbf{U}_s$ (which spans the columns space of $\hat{\mathbf{Q}}^{-\frac{1}{2}} \mathbf{A}$) can be obtained via the truncated SVD of $\overline{\mathbf{X}}$ due to the presence of noise. Hence, a more precise in terms of notation form of \eqref{eq330} is     
\begin{flalign}
\overline{\mathbf{X}} = \hat{\mathbf{U}}_{\rm s} \hat{\boldsymbol{\Sigma}}_{\rm s} \hat{\mathbf{V}}_{\rm s}^H
\label{eq356} 
\end{flalign}
where $\hat{\mathbf{U}}_{\rm s} = [\hat{\mathbf{u}}_1 \cdots \hat{\mathbf{u}}_L] \in \mathbb{C}^{M \times L }$ is the matrix of $L$ left singular vectors associated with the $L$ largest singular values on the diagonal of $\hat{\boldsymbol{\Sigma}}_{\rm s} \in \mathbb{R}^{L \times L }$. Accordingly, $\hat{\widetilde{\mathbf{U}}}_{\rm s}$ is defined as
\begin{flalign}
\hat{\widetilde{\mathbf{U}}}_{\rm s}  &\triangleq \hat{\mathbf{Q}}^{\frac{1}{2}} \hat{\mathbf{U}}_{\rm s} = \left[ (\hat{\mathbf{Q}}^{\frac{1}{2}} \hat{\mathbf{u}}_{1}) \cdots (\hat{\mathbf{Q}}^{\frac{1}{2}} \hat{\mathbf{u}}_{L}) \right] \in \mathbb{C}^{M \times L }.
\label{eq358} 
\end{flalign}

Replacing $\widetilde{\mathbf{u}}_p$ by $\hat{\widetilde{\mathbf{u}}}_p$ in \eqref{eq353} and generating $\hat{\mathbf{H}}$ and $\hat{\mathbf{h}}$, \eqref{eq353} turns into an approximate equality, i.e.,
\begin{flalign}
\hat{\mathbf{H}} \mathbf{a} \approx \hat{\mathbf{h}}.  
\label{eq359} \
\end{flalign}
The optimal value of $\mathbf{a}$ can now be obtained employing the GLS technique \cite{amemiya1985generalized}, \cite{kariya2004generalized}. 
Using \eqref{eq359}, the GLS optimization problem can be formulated as
\begin{align}
\hat{\mathbf{a}} = \mathrm{arg}\displaystyle \min_{\mathbf{a}} (\hat{\mathbf{H}} \mathbf{a}-\hat{\mathbf{h}})^H  \mathbf{W} (\hat{\mathbf{H}} \mathbf{a}-\hat{\mathbf{h}}) 
\label{eq360}
\end{align}
where $\mathbf{W} \triangleq \left( \mathbb{E}\{\hat{\mathbf{e}} \hat{\mathbf{e}}^H\} \right)^{-1}  \in \mathbb{C}^{L(|\mathcal{I}|-L) \times L(|\mathcal{I}|-L) }$ and $\hat{\mathbf{e}} \triangleq \hat{\mathbf{H}} \mathbf{a} - \hat{\mathbf{h}} \in \mathbb{C}^{L(|\mathcal{I}|-L)}$. 
The solution of \eqref{eq360} is given by
\begin{flalign}
 \hat{\mathbf{a}} = (\hat{\mathbf{H}}^H \mathbf{W} \hat{\mathbf{H}})^{-1} \hat{\mathbf{H}}^H \mathbf{W} \hat{\mathbf{h}}. 
\label{eq363} \
\end{flalign}

To use \eqref{eq363}, an estimate of $\mathbf{W}$ is required. However, it is clear from the definitions of $\mathbf{W}$ and $\hat{\mathbf{e}}$ that an estimate of $\mathbf{W}$ depends on the unknown vector $\mathbf{a}$. Thus, it is natural to utilize an iterative scheme to estimate $\hat{\mathbf{a}}$ in one step, followed by estimating $\hat{\mathbf{W}}$ in the other step by employing $\hat{\mathbf{a}}$ obtained in the previous step. The alternation between these two steps is then carried on until a termination criteria is satisfied. 

To figure out a principle for finding $\hat{\mathbf{W}}$, we take into consideration the first-order subspace estimation error by expressing $\hat{\mathbf{u}}_{p}$ as $\hat{\mathbf{u}}_{p} \triangleq \mathbf{u}_{p} + \mathbf{\Delta} \mathbf{u}_{p}$ for $p=1, \cdots, L$. Exploiting this definition, we have 
\begin{flalign}
\hat{\mathbf{e}}_p &\triangleq  \hat{\mathbf{H}}_p \mathbf{a} - \hat{\mathbf{h}}_p = \mathbf{B}^H \left(\mathbf{Z}_{\mathcal{I}} \hat{\bar{\mathbf{u}}}_p + \mathrm{diag} \{\mathbf{Z}_{\mathcal{I}} \hat{\bar{\mathbf{u}}}_p \} \mathbf{Z}_{\mathcal{I}} \mathbf{W}_{a} \mathbf{a} \right) \nonumber \\ 
 &= \mathbf{B}^H \left(\mathbf{Z}_{\mathcal{I}} \mathbf{W}_{D} \hat{\mathbf{Q}}^{\frac{1}{2}} \hat{\mathbf{u}}_{p} + \mathrm{diag} \{\mathbf{Z}_{\mathcal{I}} \mathbf{W}_{D} \hat{\mathbf{Q}}^{\frac{1}{2}} \hat{\mathbf{u}}_{p} \} \mathbf{Z}_{\mathcal{I}} \mathbf{W}_{a} \mathbf{a} \right) \nonumber \\ 
 & = \mathbf{B}^H \left(\mathbf{Z}_{\mathcal{I}} \mathbf{W}_{D} \hat{\mathbf{Q}}^{\frac{1}{2}} \hat{\mathbf{u}}_{p} + \mathrm{diag} \{\mathbf{Z}_{\mathcal{I}} \mathbf{W}_{a} \mathbf{a} \} \mathbf{Z}_{\mathcal{I}} \mathbf{W}_{D} \hat{\mathbf{Q}}^{\frac{1}{2}} \hat{\mathbf{u}}_{p} \right) \nonumber \\ 
 & = \mathbf{B}^H \left(\mathbf{I}_{\vert \mathcal{I} \vert} + \mathrm{diag} \{\mathbf{Z}_{\mathcal{I}} \mathbf{W}_{a} \mathbf{a} \} \right) \mathbf{Z}_{\mathcal{I}} \mathbf{W}_{D} \hat{\mathbf{Q}}^{\frac{1}{2}} \hat{\mathbf{u}}_{p} \nonumber \\
 & = \mathbf{C}(\mathbf{a}) \hat{\mathbf{u}}_{p} \ , \ \ \ \ p= 1, \cdots, L \
\label{eq364} \
\end{flalign}
where $\mathbf{C}(\mathbf{a}) \triangleq \mathbf{B}^H \left(\mathbf{I}_{\vert \mathcal{I} \vert} + \mathrm{diag}\{\mathbf{Z}_{\mathcal{I}} \mathbf{W}_{a} \mathbf{a} \} \right) \mathbf{Z}_{\mathcal{I}} \mathbf{W}_{D} \hat{\mathbf{Q}}^{\frac{1}{2}} \in \mathbb{C}^{(|\mathcal{I}|-L) \times M }$. In \eqref{eq364}, we also used the property $\mathrm{diag} \{ \mathbf{x}\} \mathbf{y} = \mathrm{diag} \{ \mathbf{y}\} \mathbf{x}$ as well as the equality $\hat{\bar{\mathbf{u}}}_p = \mathbf{W}_{D} \hat{\widetilde{\mathbf{u}}}_p = \mathbf{W}_{D} \hat{\mathbf{Q}}^{\frac{1}{2}} \hat{\mathbf{u}}_{p}$. 

Using the definition of $\mathbf{C} (\mathbf{a})$, \eqref{eq349} can be recast as
\begin{flalign}
\mathbf{C} (\mathbf{a}) [\mathbf{u}_{1} \mathbf{u}_{2} \cdots \mathbf{u}_{L}] = \mathbf{C} (\mathbf{a}) \mathbf{U}_{\rm s} = \mathbf{0}_{(|\mathcal{I} |-L) \times L}. 
		\label{eq366}
\end{flalign}
According to the definitions of $\mathbf{W}$, $\hat{\mathbf{e}}$ and $\mathbf{C}(\mathbf{a})$, and using \eqref{eq364} as well as \eqref{eq366}, $\hat{\mathbf{e}}$ can be found as
\begin{flalign}
	\hat{\mathbf{e}} &= \mathrm{vec} \{[\hat{\mathbf{e}}_1 \hat{\mathbf{e}}_2 \cdots \hat{\mathbf{e}}_L] \} = \mathrm{vec} \{\mathbf{C}(\mathbf{a}) \hat{\mathbf{U}}_{\rm s} \} 
	= \mathrm{vec}\{\mathbf{C}(\mathbf{a}) (\mathbf{U}_{\rm s} + \mathbf{\Delta} \mathbf{U}_{\rm s}) \} \nonumber \\ &= \mathrm{vec}\{\mathbf{C}(\mathbf{a}) \mathbf{\Delta} \mathbf{U}_{\rm s} \} 
	= (\mathbf{I}_L \otimes \mathbf{C}(\mathbf{a})) \mathrm{vec}\{\mathbf{\Delta} \mathbf{U}_{\rm s} \} = (\mathbf{I}_L \otimes \mathbf{C}(\mathbf{a})) \mathbf{\Delta} \mathbf{u}_{\rm s}
	\label{eq367} 
\end{flalign}
where $\mathbf{\Delta} \mathbf{U}_{\rm s} \triangleq [\mathbf{\Delta} \mathbf{u}_{1} \cdots \mathbf{\Delta} \mathbf{u}_{L}] \in \mathbb{C}^{M \times L}$ and $\mathbf{\Delta} \mathbf{u}_{\rm s} \triangleq \mathrm{vec} \{ \mathbf{\Delta} \mathbf{U}_{\rm s} \} \in \mathbb{C}^{ML}$. The identity $\mathrm{vec} \{\mathbf{X} \mathbf{Y} \mathbf{Z}\} = (\mathbf{Z}^T \otimes \mathbf{X}) \mathrm{vec} \{ \mathbf{Y} \}$ has also being used in \eqref{eq367}.
Moreover, according to the definitions of $\mathbf{W}$ and $\hat{\mathbf{e}}$, and using \eqref{eq367}, $\mathbf{W}$ can be expressed as
\begin{flalign}
\mathbf{W} &= \left[\left(\mathbf{I}_L \otimes \mathbf{C}(\mathbf{a})\right) \mathbb{E}\{\mathbf{\Delta} \mathbf{u}_{\rm s} \mathbf{\Delta} \mathbf{u}_{\rm s}^H\}  \left(\mathbf{I}_L \otimes \mathbf{C}(\mathbf{a})\right)^H \right]^{-1}.
\label{eq370} 
\end{flalign}

Using the first-order perturbation expansion for SVD  \cite{steinwandt2017generalized}, \cite{li1993performance}, we can write that 
\begin{flalign}
\mathbf{\Delta}\mathbf{U}_{\rm s} \approx (\mathbf{I}_M - \mathbf{U}_{\rm s} \mathbf{U}_{\rm s}^H) \overline{\mathbf{N}} \mathbf{V}_{\rm s} \boldsymbol{\Sigma}_{\rm s}^{-1}.
\label{eq371} 
\end{flalign}
Applying the vectorization operator to \eqref{eq371}, we have
\begin{flalign}
\mathbf{\Delta} \mathbf{u}_{\rm s} &= \mathrm{vec}\{ \mathbf{\Delta} \mathbf{U}_{\rm s}\} \approx \left( \boldsymbol{\Sigma}_{\rm s}^{-1} \mathbf{V}_{\rm s}^T \otimes (\mathbf{I}_M - \mathbf{U}_{\rm s} \mathbf{U}_{\rm s}^H) \right) \Bar{\mathbf{n}}  
\label{eq372} 
\end{flalign}
where $\Bar{\mathbf{n}} \triangleq \mathrm{vec} \{ \overline{\mathbf{N}} \} \in \mathbb{C}^{MN }$. Inserting \eqref{eq372} into \eqref{eq370} yields 
\begin{flalign}
\mathbf{W} &\approx [ (\mathbf{I}_L \otimes \mathbf{C}(\mathbf{a})) ( \boldsymbol{\Sigma}_{\rm s}^{-1} \mathbf{V}_{\rm s}^T \otimes (\mathbf{I}_M - \mathbf{U}_{\rm s} \mathbf{U}_{\rm s}^H) ) 
  \mathbb{E} \{\Bar{\mathbf{n}} \Bar{\mathbf{n}}^H \} (  \mathbf{V}_{\rm s}^* \boldsymbol{\Sigma}_{\rm s}^{-1} \otimes ( \mathbf{I}_M - \mathbf{U}_{\rm s} \mathbf{U}_{\rm s}^H) ) (\mathbf{I}_L \otimes \mathbf{C}^H (\mathbf{a})) ]^{-1} \nonumber \\
&= [   ( \boldsymbol{\Sigma}_{\rm s}^{-1} \mathbf{V}_{\rm s}^T \otimes \mathbf{C}(\mathbf{a}) (\mathbf{I}_M - \mathbf{U}_{\rm s} \mathbf{U}_{\rm s}^H) ) (\mathbf{I}_M \otimes \mathbf{I}_N)
 (  \mathbf{V}_{\rm s}^* \boldsymbol{\Sigma}_{\rm s}^{-1} \otimes ( \mathbf{I}_M - \mathbf{U}_{\rm s} \mathbf{U}_{\rm s}^H) \mathbf{C}^H (\mathbf{a}) ) ]^{-1}
\label{eq373} 
\end{flalign}
where $\mathbb{E} \{\Bar{\mathbf{n}} \Bar{\mathbf{n}}^H \} = \mathbf{I}_{MN} = ( \mathbf{I}_M \otimes \mathbf{I}_N )$. The property $(\mathbf{X} \otimes \mathbf{Y}) (\mathbf{Z} \otimes \mathbf{T}) = (\mathbf{X}\mathbf{Z} \otimes \mathbf{Y}\mathbf{T})$ has been used for deriving \eqref{eq373}. Using this property again together with \eqref{eq366}, \eqref{eq373} can be further simplified as
\begin{flalign}
\mathbf{W} &\approx [ ( \boldsymbol{\Sigma}_{\rm s}^{-1} \mathbf{V}_{\rm s}^T \mathbf{V}_{\rm s}^* \boldsymbol{\Sigma}_{\rm s}^{-1} \otimes \mathbf{C}(\mathbf{a}) \mathbf{C}^H (\mathbf{a})) ]^{-1} \nonumber \\
&=  ( \boldsymbol{\Sigma}_{\rm s}^{-2} \otimes \mathbf{C}(\mathbf{a}) \mathbf{C}^H (\mathbf{a})) ^{-1} = (\boldsymbol{\Sigma}_{\rm s}^{2} \otimes (\mathbf{C}(\mathbf{a}) \mathbf{C}^H (\mathbf{a}))^{-1}  ) . 
\label{eq374} 
\end{flalign}
The identities $\mathbf{V}_{\rm s}^T \mathbf{V}_{\rm s}^* = \mathbf{I}_L$ and $(\mathbf{X} \otimes \mathbf{Y})^{-1} = (\mathbf{X})^{-1} \otimes (\mathbf{Y})^{-1}$ have also been used here.

However, since the matrix $\boldsymbol{\Sigma}_{\rm s}$ is unknown, we replace it with $\hat{\boldsymbol{\Sigma}}_{\rm s}$ obtained from \eqref{eq356}. Eventually, $\hat{\mathbf{W}}$ is expressed as 
\begin{flalign}
\hat{\mathbf{W}} &\approx 
 (\hat{\boldsymbol{\Sigma}}_{\rm s}^{2} \otimes (\mathbf{C}(\mathbf{a}) \mathbf{C}^H (\mathbf{a}))^{-1}  ). 
\label{eq375} 
\end{flalign}
It can now be seen from \eqref{eq375} that $\hat{\mathbf{W}}$ is a function of $\mathbf{a}$. As a result, an iterative scheme should be used to estimate $\hat{\mathbf{a}}$ and $\hat{\mathbf{W}}$ via \eqref{eq363} and \eqref{eq375} in alternative manner. The LS solution of \eqref{eq359} can be used as the initial vector for $\hat{\mathbf{a}}$, i.e., 
\begin{flalign}
\hat{\mathbf{a}}^{(0)} \triangleq \hat{\mathbf{a}}_{LS} = \hat{\mathbf{H}}^{\dagger} \hat{\mathbf{h}}.
\label{eq376}
\end{flalign}
Initializing $\mathbf{C}(\mathbf{a})$ by inserting \eqref{eq376} into the definition of $\mathbf{C}(\mathbf{a})$, enables us to obtain
\begin{flalign}
\hat{\mathbf{W}}^{(0)} = 
 (\hat{\boldsymbol{\Sigma}}_{\rm s}^{2} \otimes (\mathbf{C}(\hat{\mathbf{a}}^{(0)}) \mathbf{C}(\hat{\mathbf{a}}^{(0)})^H)^{-1}  ). 
\label{eq377}
\end{flalign}
Subsequently, a new estimate of $\hat{\mathbf{a}}$ is generated by substituting \eqref{eq377} into \eqref{eq363}. The iterations carry on until a proper termination criteria is satisfied\footnote{Although any common termination criteria can be adopted, we observe that performing 3 to 5 iterations are usually sufficient to obtain a precise result. Therefore, in the numerical examples, 5 is opted as the number of iterations for implementing the proposed methods.}. 

The remaining problem is still how to determine the members of $\mathcal{I}$, i.e., the indices of the selected equations of \eqref{eq345} to be employed for estimating DOAs. Given the cardinality $|\mathcal{I}|$, it is reasonable to select those indices which correspond to the entries of $\hat{\Bar{\mathbf{u}}}_1 = \mathrm{DFT} \{ \hat{\widetilde{\mathbf{u}}}_1 \}$ with $|\mathcal{I}|$ largest absolute values. According to \eqref{eq358}, $\hat{\widetilde{\mathbf{u}}}_1 = \hat{\mathbf{Q}}^{\frac{1}{2}} \hat{\mathbf{u}}_{1}$ with $\hat{\mathbf{u}}_{1}$ denoting the left singular vector of $\overline{\mathbf{X}}$ which corresponds to the largest singular value. The logic of this choice is rooted in \eqref{eq332}, where it is indicated that each $\hat{\widetilde{\mathbf{u}}}_p$ can be expressed as a linear combination of the columns of $\mathbf{A}$. Therefore, picking the indices of $\hat{\Bar{\mathbf{u}}}_1 $ with largest absolute values is a sensible choice because of the following three reasons. 1)~The structure of the DFT basis is completely matched with the columns of $\mathbf{A}$, and consequently $\hat{\widetilde{\mathbf{u}}}_p$'s, which makes the absolute values of $\hat{\Bar{\mathbf{u}}}_p $'s the best option to be used for selecting the most relevant equations. 2)~Choosing the indices with largest absolute values guarantees picking equations with the most contributions. 3)~The estimation error of finding $\hat{\mathbf{u}}_{1}$ is the smallest among $\hat{\mathbf{u}}_{p}$'s since it associates with the largest singular values, resulting in smaller error. Finally, the steps required for implementing the proposed DOA estimation method are summarized in Algorithm~2.

\begin{table}[!h]
	\label{AL2}
		\begin{tabular}{l}
			\hline
			\textbf{Algorithm~2:} Forward-only DOA Estimation 
		\\ \hline
		1: Compute $\hat{\mathbf{R}} =  1 / N \sum_{t=1}^{N} \mathbf{x}(t) \mathbf{x}^H(t) $ and estimate $\hat{\mathbf{Q}}$ using Algorithm~{1}. \\
		2: Calculate $\overline{\mathbf{X}} = \hat{\mathbf{Q}}^{-\frac{1}{2}
		} \mathbf{X} $ and construct $\mathbf{W}_{a}$ as well as $\overline{{\mathbf{W}}}$. \\
		3: For a pre-chosen $|\mathcal{I}|$, determine $\mathbf{Z}_{\mathcal{I}}$ and obtain $\mathbf{B}$ via performing the SVD of the matrix $\widetilde{{\mathbf{Z}}}_{\mathcal{I}} = \mathbf{Z}_{\mathcal{I}} \overline{\mathbf{W}}$ so that the \\ condition $\mathbf{B}^H \widetilde{{\mathbf{Z}}}_{\mathcal{I}} = \mathbf{0}$ is  satisfied. \\
		4: Carry out the SVD of $\overline{\mathbf{X}}$ to obtain $\hat{\mathbf{U}}_{\rm s}$ and $\hat{\boldsymbol{\Sigma}}_{\rm s}$, where the former contains the $L$ left singular vectors corresponding to \\ the $L$ largest singular values on the diagonal of the latter. \\
		5: Utilize \eqref{eq358} for computing $\hat{\widetilde{\mathbf{U}}}_{\rm s}$. Apply DFT on the columns of $\hat{\widetilde{\mathbf{U}}}_{\rm s}$ to obtain $\hat{\bar{\mathbf{u}}}_p$'s. \\
		6: Form $\hat{\mathbf{H}}$ and $\hat{\mathbf{h}}$ using the estimates $\hat{\bar{\mathbf{u}}}_p$'s instead of $\bar{\mathbf{u}}_p$'s. \\ 
		7: Set $i=0$ and $\hat{\mathbf{a}}^{(0)} = \hat{\mathbf{H}}^{\dagger} \hat{\mathbf{h}}$. In addition, set the maximum number of iterations $i_{max} = 5 $.  \\
		\textbf{while} $ i \leq i_{max}$ \\
		8: Compute $\mathbf{C} \left( \hat{\mathbf{a}}^{(i)} \right)$ and $\hat{\mathbf{W}}^{(i)}$. 
		\\
		9: Generate a new estimate $\hat{\mathbf{a}}^{(i+1)} = \left(\hat{\mathbf{H}}^H \hat{\mathbf{W}}^{(i)} \hat{\mathbf{H}} \right)^{-1} \hat{\mathbf{H}}^H \hat{\mathbf{W}}^{(i)} \hat{\mathbf{h}}$. \\
		10: set $ i=i+1$. \\
		\textbf{end} \\
		11: Find the $L$ roots of the polynomial $\gamma^L + \sum_{l=1}^{L} [\hat{\mathbf{a}}]_l \ \gamma^{L-l} = 0$, denoted by $\hat{\gamma}_l, l=1, \cdots, L$. \\
		12: Obtain the $L$ DOA estimates as $\hat{\theta}_l = \mathrm{arcsin} \left( -\frac{\beta_l \lambda}{2 \pi d} \right)$, where $\beta_l$ is the phase argument of $\hat{\gamma}_l, \; l=1, \cdots, L$.\\
		\hline
		\end{tabular}
 \end{table}

To provide a theoretical measure for performance of the proposed forward-only DOA estimation method, the asymptotic variance of the $l$th DOA estimated by the proposed forward-only algorithm is derived under a high SNR assumption in the following proposition.   

\textit{Proposition 2:} The asymptotic variance of the proposed forward-only DOA estimation algorithm, for a particular matrix $\mathbf{Z}_{\mathcal{I}}$, is given as
\begin{flalign}
	\mathbb{E} \{ \Delta \theta_l^2 \} &\approx  \frac{1}{2} \left(\frac{ \lambda}{2 \pi d \cos(\theta_l)}\right)^2 \frac{\boldsymbol{\gamma}_l^T (\mathbf{H}^H \mathbf{W} \mathbf{H})^{-1} \boldsymbol{\gamma}_l^*}{| \phi_l|^2}
	\label{eq396} 
\end{flalign}
where $\boldsymbol{\gamma}_l \triangleq [\gamma_l^{L-1} \cdots 1]^T$ and $\phi_l \triangleq L \gamma_l^{L-1} + (L-1) [\mathbf{a}]_1 \gamma_l^{L-2} + \cdots + [\mathbf{a}]_{L-1}$. 

\textit{Proof:} See Appendix~\ref{APP3}. \ \  \ \ \  \ \ \ \ \ \  \ \ \  \ \ \  \ \ \ \ \ \  \ \ \  \ \ \  \ \ \ \ \ \  \ \ \  \ \ \  \ \ \ \ \ \  \ \ \  \ \ \  \ \ \ \ \ \  \ \ \  \ \ \  \ \ \ \ \ \  \ \ \ \ \ \  $\blacksquare$

We aim to generate double number of DOA candidates\footnote{In fact, the number of DOA candidates can be arbitrary, but from diverse numerical simulations conducted, we find that generating more candidates than $2L$ does not improve the DOA estimation accuracy considerably.} by running the proposed Algorithm~2 twice for two different values of $|\mathcal{I}|$. Then a proper DOA selection strategy should be employed to determine the final DOA estimates. First, consider the following example to get a more through wisdom about how different choices of $|\mathcal{I} |$ affect the DOA estimation accuracy. 

\begin{figure}[ht]
	\begin{center}
		{\includegraphics[width=3.7in]{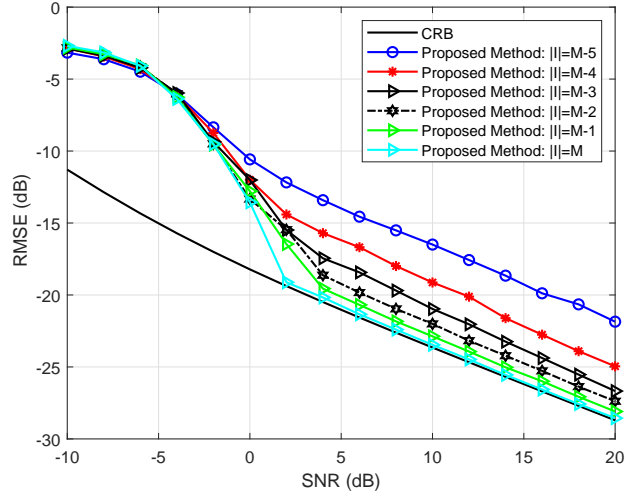}}
		\caption{RMSE performance of the proposed method for different $| \mathcal{I} |$ vs. SNR for $L=2$ uncorrelated sources with $\boldsymbol{\theta} = [-2^{\circ}  ,  7^{\circ} ]$, $M=8$, and $N=40$.}
		\label{fig1}
	\end{center}
\end{figure}

\textit{Illustrative Example 1:} Consider a scenario where the signals of two uncorrelated sources located in $\boldsymbol{\theta} = [-2^{\circ}, 7^{\circ}]$ are received by a ULA consisting $M=8$ sensors with half waveform adjacent distances. The sensor noise covariance matrix is set as $\mathbf{Q} = \mathrm{diag} \{[10, 1.2, 3.5, 18, 2, 8.5, 24, 6.5 ] \}$, the sample size is $N=40$, and 2000 Monte Carlo runs are conducted to calculate RMSE defined as
\begin{flalign}
{\rm RMSE} = 10 {\rm log}_{10}\sqrt{\frac{1}{2000L}\displaystyle\sum_{l=1}^{L} \displaystyle\sum_{i=1}^{2000} (\hat{\theta}_{l,i}-\theta_l)^2}.
\label{eq378}
\end{flalign}
We also include the deterministic CRB \cite{pesavento2001maximum} as a benchmark. In Fig.~\ref{fig1}, higher accuracy in DOA estimation can be observed for larger $|\mathcal{I}|$. From this observation, we conclude that $|I|=M-1$ and $|I|=M$ are the best choices in the sense of providing the most precise estimates for generating double number of DOA candidates.  

\subsection{DOA Selection Strategy}
After running Algorithm~2 twice with $| \mathcal{I}|= M-1$ and $| \mathcal{I}|= M$, $2L$ DOA candidates are generated. The natural question is that how to select $L$ final DOA estimates. Two known conventional approaches to DOA selection are based on conventional beamformer (CB) \cite{qian2013improved}, \cite{gershman1998pseudo, vasylyshyn2012improved, vasylyshyn2013removing} and ML cost function minimization \cite{shaghaghi2015subspace}, \cite{qian2016enhanced}, \cite{gershman1999new}. Recently, another method, which has been originally proposed for joint source number detection and DOA estimation \cite{izedi2017joint}, has been employed as the DOA selection scheme. It is based on the generalized likelihood ratio (GLR), which extracts the final $L$ DOAs sequentially \cite{esfandiari2021enhanced}. The computational complexity of such DOA selection strategy is much lower than that of based on the ML cost function minimization. Besides, the performance provided by the GLR is comparable with that provided by the ML-based methods. Here we aim to design a DOA selection strategy which takes advantage of the three aforementioned approaches, i.e, it uses CB, deterministic ML cost function, and the GLR technique. 

In doing so, the following three-step selection strategy is proposed. 

\textit{Step 1}: 
Denote the vector containing all $2L$ DOA candidates by $\boldsymbol{\theta}_{2L}$. Calculate the CB output for a proper number of equidistant points to cover the whole interval of interest, i.e., $[-\frac{\pi}{2}, \frac{\pi}{2}]$. Then, find the $(L+1)$th peak and define a threshold, denoted by $\eta$, as the CB output at the $(L+1)$th peak\footnote{Note that if the total number of peaks is smaller than $(L+1)$, select the last peak with the lowest output.}. Afterwards, calculate the CB output for entries of $\boldsymbol{\theta}_{2L}$ and stack those entries with output larger than $\eta$ in a new vector $\widetilde{\boldsymbol{\theta}}$. If the number of entries in $\widetilde{\boldsymbol{\theta}}$ is less than $L$, then let $\widetilde{\boldsymbol{\theta}} = \boldsymbol{\theta}_{2L}$.   

\textit{Step 2}: Determine the first DOA as that entry of $\widetilde{\boldsymbol{\theta}}$ which maximizes the GLR, i.e.,
\begin{flalign}
\hat{\theta}_1 = \mathrm{arg}\displaystyle \max_{\theta} \frac{\mathbf{a}^H (\theta) \hat{\mathbf{Q}}^{-1} \hat{\mathbf{R}} \hat{\mathbf{Q}}^{-1}  \mathbf{a} (\theta)}{\mathbf{a}^H (\theta) \hat{\mathbf{Q}}^{-1}  \mathbf{a} (\theta)} \ \ \ , \ \theta \in \widetilde{\boldsymbol{\theta}}.
\label{eq479}
\end{flalign}
The GLR in \eqref{eq479} is a straightforward nonuniform noise extension of the GLR in \cite{izedi2017joint} for the case of uniform noise. 

\textit{Step 3}: Denote the remaining entries of $\widetilde{\boldsymbol{\theta}}$ as $\bar{\boldsymbol{\theta}}$, and the size of $\bar{\boldsymbol{\theta}}$ as $\bar{L}$. Then, divide the $\bar{L}$ DOA candidates in $\bar{\boldsymbol{\theta}}$ into $\bar{G} = \frac{\bar{L}}{(L-1)!(\bar{L}-L+1)!}$ subsets containing $(L-1)$ different DOAs each. Denote these subsets as $\Theta_1, \cdots, \Theta_{\bar{G}}$ and associate them with $\mathbf{A}(\Theta_1), \cdots, \mathbf{A}(\Theta_{\bar{G}})$. The subset that minimizes the following deterministic ML cost function\footnote{Since the difference of employing the deterministic ML and stochastic ML is marginal, the deterministic ML is discussed here because it also has lower complexity.} determines the $(L-1)$ remaining DOAs
\begin{flalign}
\hat{\Theta}_R = \mathrm{arg}\displaystyle \min_{\Theta_{\mathbb{S}}} & \ \mathrm{trace} \left[ \left(\mathbf{P}^{\perp}_{\widetilde{A}(\Theta_{\mathbb{S}})} - \boldsymbol{\nu}_1 \boldsymbol{\nu}_1^H \right) \hat{\mathbf{Q}}^{-\frac{1}{2}} \hat{\mathbf{R}} \hat{\mathbf{Q}}^{-\frac{1}{2}} \right], \ \ \
\mathbb{S} \in \{1, \cdots, \bar{G}   \}
\label{eq469}
\end{flalign}
where $\mathbf{P}^{\perp}_{\widetilde{A}(\Theta_{\mathbb{S}})} \triangleq \mathbf{I}_M - \widetilde{\mathbf{A}}(\Theta_{\mathbb{S}}) \left( \widetilde{\mathbf{A}}(\Theta_{\mathbb{S}})^H \widetilde{\mathbf{A}}(\Theta_{\mathbb{S}}) \right)^{-1} \widetilde{\mathbf{A}}(\Theta_{\mathbb{S}})^H$, $\widetilde{\mathbf{A}}(\Theta_{\mathbb{S}}) \triangleq \hat{\mathbf{Q}}^{-\frac{1}{2}} \mathbf{A}(\Theta_{\mathbb{S}})$, and 
$\boldsymbol{\nu}_1 \triangleq \frac{\mathbf{P}^{\perp}_{\widetilde{A}(\Theta_{\mathbb{S}})} \hat{\mathbf{Q}}^{-\frac{1}{2}} \mathbf{a}(\hat{\theta}_1)}{\|  \mathbf{P}^{\perp}_{\widetilde{A}(\Theta_{\mathbb{S}})} \hat{\mathbf{Q}}^{-\frac{1}{2}} \mathbf{a}(\hat{\theta}_1) \|_2}$. 
In \eqref{eq469}, the contribution of previously estimated $\hat{\theta}_1$ on the deterministic ML cost function is isolated in $\boldsymbol{\nu}_1$ thanks to the properties of the orthogonal projection matrix \cite{kay1993fundamentals}. In addition, $ \hat{\mathbf{Q}}^{-\frac{1}{2}} \hat{\mathbf{R}} \hat{\mathbf{Q}}^{-\frac{1}{2}}$ is used instead of $ \hat{\mathbf{R}}$ to consider the general case of nonuniform noise. 

Finally, $\hat{\theta}_1$ and $\hat{\Theta}_R$ obtained via \eqref{eq469} form together the final $L$ DOA estimates.

\subsection{Computational Complexity}
For implementing the proposed DOA estimation method, Algorithm~1 should be implemented first that requires $\mathcal{O} \left( M^2 N \right)$ flops for calculating $\hat{\mathbf{R}}$ and $\mathcal{O} \left( M^3 \right)$ flops for computing GED of the pair of matrices $\left\{ \hat{\mathbf{R}}, \; \hat{\mathbf{Q}}^{(i)} \right\}$ in the $i$th iteration. Thus, the total computational complexity of Algorithm~{1} is $\mathcal{O} \left(I_1(M^3)+ M^2 N \right)$ with $I_1$ being the number of iterations. The computational complexity of performing SVD of $\overline{\mathbf{X}}$ is $\mathcal{O} \left( \mathrm{max}(M,N) \mathrm{min}(M,N)^2 \right)$, and the computational complexity of computing $\mathbf{B}$ is $\mathcal{O} \left( | \mathcal{I}| L^2 \right)$. Applying DFT on the columns of $\hat{\widetilde{U}}_{\rm s}$ requires $\mathcal{O} \left( LM \mathrm{log}_2 (M) \right)$ flops. In addition, the main source of computational complexity for calculating $\hat{\mathbf{W}}$ is the inversion of the matrix $\mathbf{C}(\mathbf{a}) \mathbf{C}^H (\mathbf{a})$, which involves $\mathcal{O} \left( (|\mathcal{I}|-L)^3 \right)$ flops, making the computational complexity of computing $\hat{\mathbf{W}}$ be $\mathcal{O} \left( L(|\mathcal{I}|-L)^3 \right)$. Determining $\hat{\mathbf{a}}$ requires $\mathcal{O} \left( L^3( 2(M- L)+1) +  L^2 (2(M- L)^2 +(M-L)) \right)$ flops. As the proposed DOA estimation algorithm is run twice with $|\mathcal{I}|=M-1$ and $|\mathcal{I}|=M$, by considering $|\mathcal{I} | \approx M$, the total complexity required to generate $2L$ DOA candidates is $ \mathcal{O} \left( 
M^2N + I_1(M^3) + (\mathrm{max}(M,N) \mathrm{min}(M,N)^2) + ML^2 +\right.$ $\left. LM \mathrm{log}_2 (M)+ I_2 ( M^3 L + M L^3 -L^4 ) \right)$ where $I_2$ denotes the number of iterations needed for finding the best $\hat{\mathbf{a}}$. At last, the complexity of DOA selection is mainly in \textit{step~3}, which is about $\mathcal{O} \left( \bar{G} (M^3 + 3M(L-1)^2 + (L-1)^3) \right)$. Finally, the total complexity of DOA estimation is reduced to $\mathcal{O} \left( M^2 N + \bar{G} M^3 \right)$ in the case that $\bar{G} \gg \mathrm{max}(I_1,I_2)$ and $M \gg L$. The significant point here is that $\bar{G}$ is approximately four times smaller than the parameter $G$ in \cite{qian2016enhanced}, resulting in the computational complexity of our method being approximately a quarter of what is required for implementing the EPUMA.

\subsection{FBA Extension}
FBA is a natural extension/improvement of the forward-only-based DOA estimation methods \cite{haardt1995unitary}, \cite{huarng1991unitary, linebarger1994efficient, gershman1999unitary, rao1993weighted, wen2017spatial}. The essence of FBA is to first transform the observed signal or SCM into a new centro-Hermitian signal matrix or centro-Hermition covariance matrix, respectively, followed by computationally simplified and more accurate DOA estimation. Higher DOA estimation accuracy is a consequence of decorrelating possibly correlated source pairs and obtaining more accurate SCM estimation.

The FBA covariance matrix is given as \cite{pesavento2000unitary}
\begin{flalign}
\mathbf{R}_{\rm FB} &= \frac{1}{2} \left(\mathbf{R} + \mathbf{J}_M \mathbf{R}^* \mathbf{J}_M   \right)  
= \frac{1}{2} \left( \mathbf{A} \mathbf{P} \mathbf{A}^H + \mathbf{Q} + \mathbf{J}_M (\mathbf{A}^* \mathbf{P}^* \mathbf{A}^T + \mathbf{Q}) \mathbf{J}_M \right) .
\label{eq379}
\end{flalign}
It can be readily verified that $\mathbf{R}_{\rm FB}$ is centro-Hermitian, i.e., $\mathbf{R}_{\rm FB}= \mathbf{J}_M \mathbf{R}^*_{FB} \mathbf{J}_M$. Rearranging \eqref{eq379}, we have  
\begin{flalign}
\mathbf{R}_{FB} &= \frac{1}{2} \left( \mathbf{A} \mathbf{P} \mathbf{A}^H + \mathbf{Q} +  \mathbf{A} \mathbf{D} \mathbf{P}^* \mathbf{D}^H \mathbf{A}^H + \mathbf{J}_M \mathbf{Q} \mathbf{J}_M \right) 
 = \mathbf{A} \widetilde{\mathbf{P}} \mathbf{A}^H +  \frac{1}{2} \widetilde{\mathbf{Q}}  
\label{eq380}
\end{flalign}
where $\widetilde{\mathbf{P}} \triangleq \frac{1}{2}( \mathbf{P} + \mathbf{D} \mathbf{P}^* \mathbf{D}^H)$, $\widetilde{\mathbf{Q}} \triangleq \mathbf{Q} + \mathbf{J}_M \mathbf{Q} \mathbf{J}_M$, $\mathbf{D} \triangleq \mathrm{diag} \left\{ e^{-j(2 \pi / \lambda) d (M-1) \sin (\theta_1)}, \cdots , e^{-j(2 \pi / \lambda) d (M-1) \sin (\theta_L)} \right\}$.
	
According to \eqref{eq380}, a proper centro-Hermitian matrix similar to $\overline{\mathbf{X}}$ defined in \eqref{eq329}, can be formed as 
\begin{flalign}
\overline{\mathbf{X}}_{\rm FB} &=  \left[  \hat{\widetilde{\mathbf{Q}}}^{-\frac{1}{2}} \mathbf{X} \ \ \ \   \mathbf{J}_M \hat{\widetilde{\mathbf{Q}}}^{-\frac{1}{2}} \mathbf{X}^* \mathbf{J}_N \right]
\label{eq384}
\end{flalign}
where $ \hat{\widetilde{\mathbf{Q}}} \triangleq \hat{\mathbf{Q}} +   \mathbf{J}_M \hat{\mathbf{Q}} \mathbf{J}_M $ is used instead of $\widetilde{\mathbf{Q}}$. The matrix $\overline{\mathbf{X}}_{\rm FB}$ can be decomposed by applying the truncated SVD as
\begin{flalign}
\overline{\mathbf{X}}_{FB} =  \hat{\mathbf{E}}_{\rm s} 
       \hat{\boldsymbol{\Pi}}_{\rm s} \hat{\mathbf{T}}_{\rm s}^H         
\label{eq385} 
\end{flalign}
where $\hat{\mathbf{E}}_{\rm s}  = [\hat{\mathbf{e}}_{1} \cdots \hat{\mathbf{e}}_{L}] \in \mathbb{C}^{M \times L }$ is composed of $L$ left singular vectors associated with $L$ largest singular values on the diagonal of $\hat{\boldsymbol{\Pi}}_{\rm s} \in \mathbb{R}^{L \times L}$. The columns of $\hat{\mathbf{E}}_{\rm s}$ and the columns of $\hat{\widetilde{\mathbf{Q}}}^{-\frac{1}{2}}\mathbf{A}$ span the same vector space. Thus, similar to the forward-only case, the relationship between $\hat{\mathbf{E}}_{\rm s}$ and $\hat{\widetilde{\mathbf{Q}}}^{-\frac{1}{2}}\mathbf{A}$ can be written as $\hat{\mathbf{E}}_{\rm s} = \hat{\widetilde{\mathbf{Q}}}^{-\frac{1}{2}}\mathbf{A} \overline{\mathbf{G}}$. Multiplying both sides of the latter equation by $\hat{\widetilde{\mathbf{Q}}}^{\frac{1}{2}}$, we obtain the FBA analog of \eqref{eq332}, that is, 
\begin{flalign}
\hat{\widetilde{\mathbf{E}}}_{\rm s} \triangleq \hat{\widetilde{\mathbf{Q}}}^{\frac{1}{2}} \hat{\mathbf{E}}_{\rm s} = \mathbf{A} \overline{\mathbf{G}}.
\label{eq487}
\end{flalign}

Following the same steps as in Subsection~\ref{sub31}, the DFT of the columns of $\hat{\widetilde{\mathbf{E}}}_{\rm s}$ can be found as 
\begin{flalign}
\hat{\overline{\mathbf{E}}}_{\rm s}  &= [\hat{\Bar{\mathbf{e}}}_{1} \cdots \hat{\Bar{\mathbf{e}}}_{L}] \triangleq \mathrm{DFT} \{ \hat{\widetilde{\mathbf{E}}}_{\rm s} \} = \mathrm{DFT} \{ \hat{\widetilde{\mathbf{Q}}}^{\frac{1}{2}} \hat{\mathbf{E}}_{\rm s} \} \nonumber \\ 
&= \left[ (\mathbf{W}_D \hat{\widetilde{\mathbf{Q}}}^{\frac{1}{2}} \hat{\mathbf{e}}_{1}) \cdots (\mathbf{W}_D \hat{\widetilde{\mathbf{Q}}}^{\frac{1}{2}} \hat{\mathbf{e}}_{L}) \right] \in \mathbb{C}^{M \times L }. \  
\label{eq387} 
\end{flalign}
Considering a specific value for $|\mathcal{I}|$, a system of linear equations similar to \eqref{eq359} can be formulated as
\begin{flalign}
\hat{\widetilde{\mathbf{H}}} &\mathbf{a} \approx  \hat{\widetilde{\mathbf{h}}} 
\label{eq392}
\end{flalign}
where $\hat{\widetilde{\mathbf{H}}}_p = \mathbf{B}^H \mathrm{diag} (\mathbf{Z}_{\mathcal{I}} \hat{\bar{\mathbf{e}}}_p) \mathbf{Z}_{\mathcal{I}} \mathbf{W}_{a} \in \mathbb{C}^{(|\mathcal{I}|-L) \times L }$, $\hat{\widetilde{\mathbf{h}}}_p = -\mathbf{B}^H \mathbf{Z}_{\mathcal{I}} \hat{\bar{\mathbf{e}}}_p \in \mathbb{C}^{(|\mathcal{I}|-L) }$, $\hat{\widetilde{\mathbf{H}}} = \left[\hat{\widetilde{\mathbf{H}}}_1^T \cdots \hat{\widetilde{\mathbf{H}}}_L^T \right]^T \in \mathbb{C}^{L(|\mathcal{I}|-L) \times L }$, $\hat{\widetilde{\mathbf{h}}} = \left[ \hat{\widetilde{\mathbf{h}}}_1^T \cdots \hat{\widetilde{\mathbf{h}}}_L^T \right]^T \in \mathbb{C}^{L(|\mathcal{I}|-L)}$ with the entries of $\mathbf{a}$ being the coefficients of the polynomial presented in \eqref{eq343}. Finally, the GLS solution of \eqref{eq392} is given by 
\begin{flalign}
 \hat{\mathbf{a}} = \left( \hat{\widetilde{\mathbf{H}}}^H \hat{\mathbf{W}}_{FB} \hat{\widetilde{\mathbf{H}}} \right)^{-1} \hat{\widetilde{\mathbf{H}}}^H \hat{\mathbf{W}}_{FB} \hat{\widetilde{\mathbf{h}}}  
\label{eq393} 
\end{flalign}
where $\hat{\mathbf{W}}_{\rm FB} \approx \hat{\boldsymbol{\Pi}}_{\rm s}^{2} \otimes \left(\mathbf{C}_{\rm FB} (\mathbf{a}) \mathbf{C}_{\rm FB}(\mathbf{a})^H \right)^{-1}$, $\mathbf{C}_{\rm FB}(\mathbf{a}) \triangleq \mathbf{B}^H \left(\mathbf{I}_{\vert \mathcal{I} \vert} + \mathrm{diag} \{\mathbf{Z}_{\mathcal{I}} \mathbf{W}_{a} \mathbf{a} \} \right) \mathbf{Z}_{\mathcal{I}} \mathbf{W}_{D} \hat{\widetilde{\mathbf{Q}}}^{\frac{1}{2}} \in \mathbb{C}^{(|\mathcal{I}|-L) \times M }$. 

After finding $\hat{\mathbf{a}}$ using \eqref{eq393}, the $L$ DOA estimates $\hat{\theta}_l$, $l=1, \cdots, L$ are obtained as $\hat{\theta}_l = \mathrm{arcsin} \left( -\frac{\beta_l \lambda}{2 \pi d} \right)$ with $\beta_l$ denoting the phase argument of $\hat{\gamma}_l$. In addition, $\hat{\gamma}_l$ denotes the $l$th root of the polynomial defined as $\gamma^L + \sum_{l=1}^{L} [\hat{\mathbf{a}}]_l \ \gamma^{L-l} = 0$.

\textbf{Remark 4}: Similar to \textit{Proposition~2}, it can be readily shown that the asymptotic variance of the DOA estimation for the FBA extension of the proposed method in the high SNR regime is given as
\begin{flalign}
\mathbb{E} \{ \Delta \theta_l^2 \} &\approx  \frac{1}{2} \left(\frac{ \lambda}{2 \pi d \cos(\theta_l)}\right)^2 \frac{\boldsymbol{\gamma}_l^T \left(\widetilde{\mathbf{H}}^H \mathbf{W}_{\rm FB} \widetilde{\mathbf{H}} \right)^{-1} \boldsymbol{\gamma}_l^*}{| \phi_l|^2}.
\label{eq399} 
\end{flalign}

\section{SIMULATION RESULTS}
\label{sec5}
The aim of this section is to evaluate the performance of the proposed method and compare it to that of the state-of-the-art algorithms in terms of diverse numerical simulation examples especially for challenging scenarios. Our examples address  both uniform and nonuniform sensor noise cases. For the uniform noise case, the performance of the forward-only and FBA versions of the proposed method is compared with that of the unitary root-MUSIC method \cite{pesavento2000unitary}, the root-swap unitary root-MUSIC method \cite{shaghaghi2015subspace}, the EPUMA method \cite{qian2016enhanced}, and the UE GLS \cite{steinwandt2017generalized}. For the nonuniform noise case, the ``NISB+MUSIC'' method \cite{esfandiari2019non} and the ``IMLSE+MUSIC'' method \cite{liao2016iterative} are used for comparison. In addition, for achieving better DOA estimations, the combinations of both the NISB and IMLSE with the root-MUSIC framework are considered for the nonuniform noise case. The uniform stochastic CRB \cite{stoica1990performance}, and the nonuniform stochastic CRB \cite{pesavento2001maximum} are used as the benchmarks in the corresponding examples. The number of trials used for calculating the RMSE is 2000 in all examples. If not further specified, a ULA with $M=10$ sensors separated by half wavelength collecting $N=10$ snapshots is considered for the uniform noise examples, while $M=8$ for the nonuniform noise examples. The SNR is computed as SNR~$ = \frac{1}{\sigma^2}$ for the uniform noise case, and as SNR~$= \frac{\sigma_{\rm s}^2}{M} \sum_{m=1}^M \frac{1}{\sigma_m^2}$ for the nonuniform noise case. Here the powers of different sources are considered to be identical and denoted by $\sigma_{\rm s}^2$. 

\begin{figure}[!]
	\begin{center}
		{\includegraphics[width=3.5in]{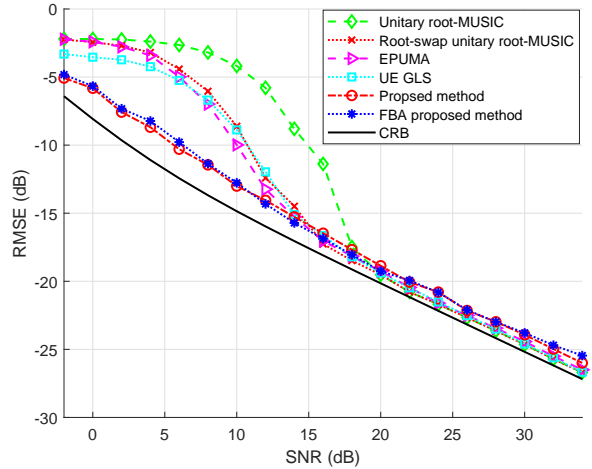}}
		\vspace{-0.5cm}
		\caption{RMSE vs. SNR for $L=3$ uncorrelated sources with $\boldsymbol{\theta} = [19^{\circ},  34^{\circ}, 36^{\circ}]$, $M=10$, and $N=10$. }
		\label{fig2}
	\end{center}
\end{figure}

\begin{figure}[!]
	\begin{center}
		{\includegraphics[width=3.5in]{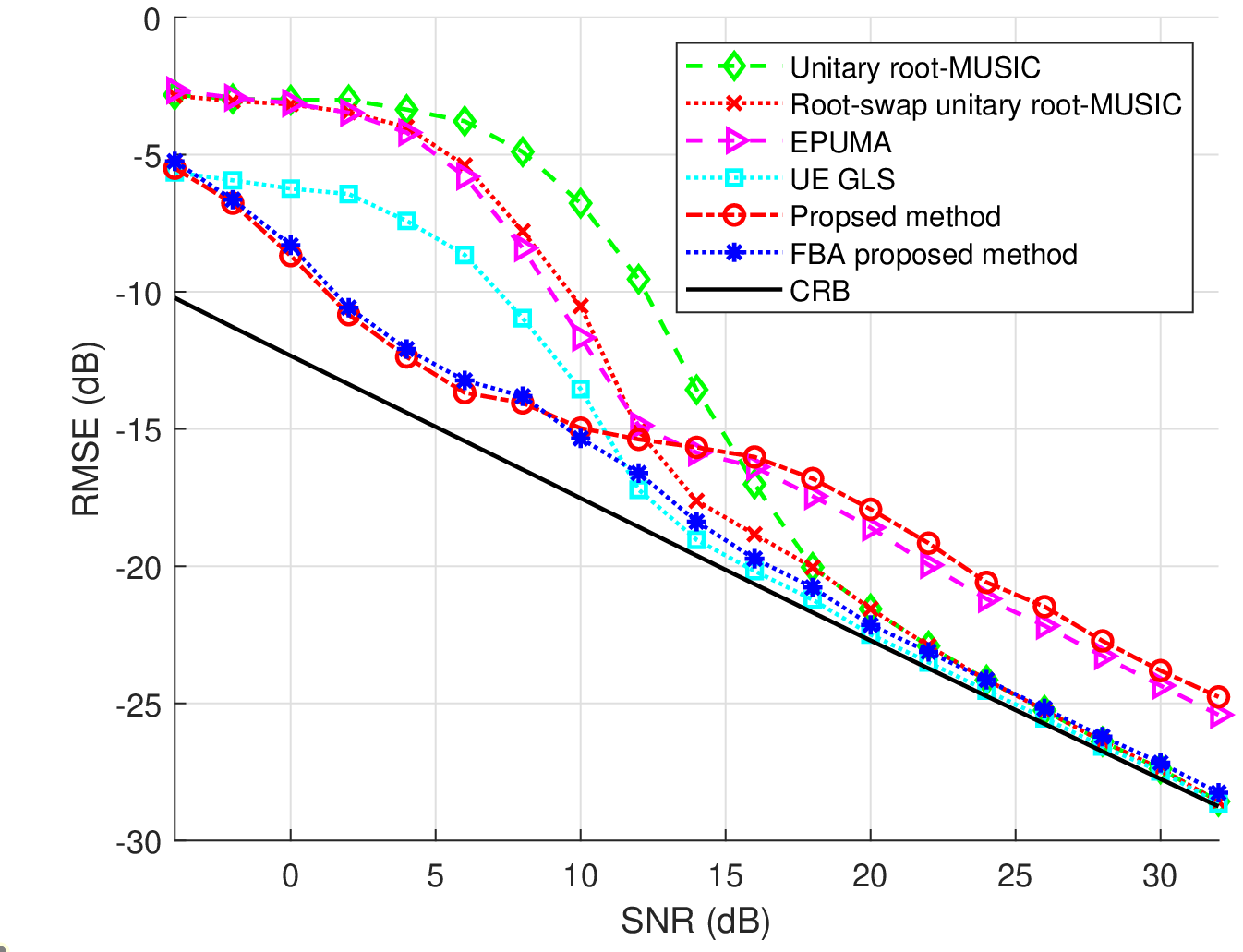}}
		\vspace{-0.5cm}
		\caption{RMSE vs. SNR for $L=2$ correlated sources with $\boldsymbol{\theta} = [5^{\circ}, 8^{\circ}]$, $\rho=0.95$, $M=10$, and $N=10$.}
		\label{fig3}
	\end{center}
\end{figure}

\begin{figure}[!]
	\begin{center}
		{\includegraphics[width=3.5in]{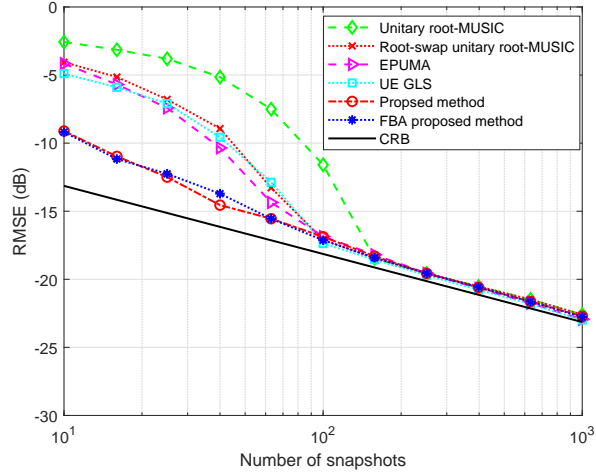}}
		\vspace{-0.5cm}
		\caption{RMSE vs. the number of snapshots for $L=3$ uncorrelated sources with $\boldsymbol{\theta} = [10^{\circ}, 34^{\circ}, 36^{\circ}]$, SNR~$ =5$~dB, and $M=10$.}
		\label{fig4}
	\end{center}
\end{figure}

\begin{figure}[!]
	\begin{center}
		{\includegraphics[width=3.5in]{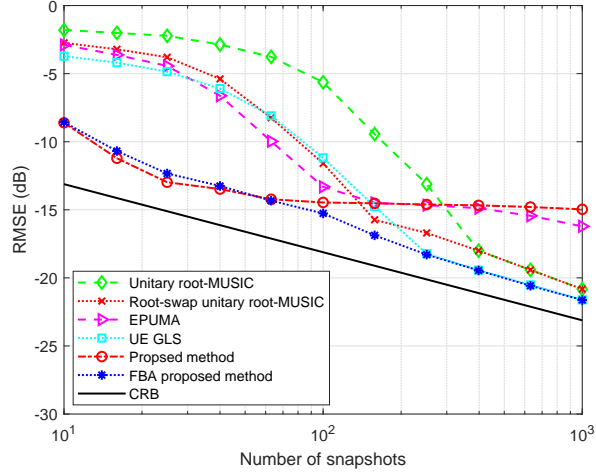}}
		\vspace{-0.5cm}
		\caption{RMSE vs. the number of snapshots for $L=2$ correlated sources with $\boldsymbol{\theta} = [ 34^{\circ}, 38^{\circ}]$, $\rho=0.95$, SNR~$ =2$~dB, and $M=10$.}
		\label{fig5}
	\end{center}
\end{figure}

\begin{figure}[!]
	\begin{center}
		{\includegraphics[width=3.5in]{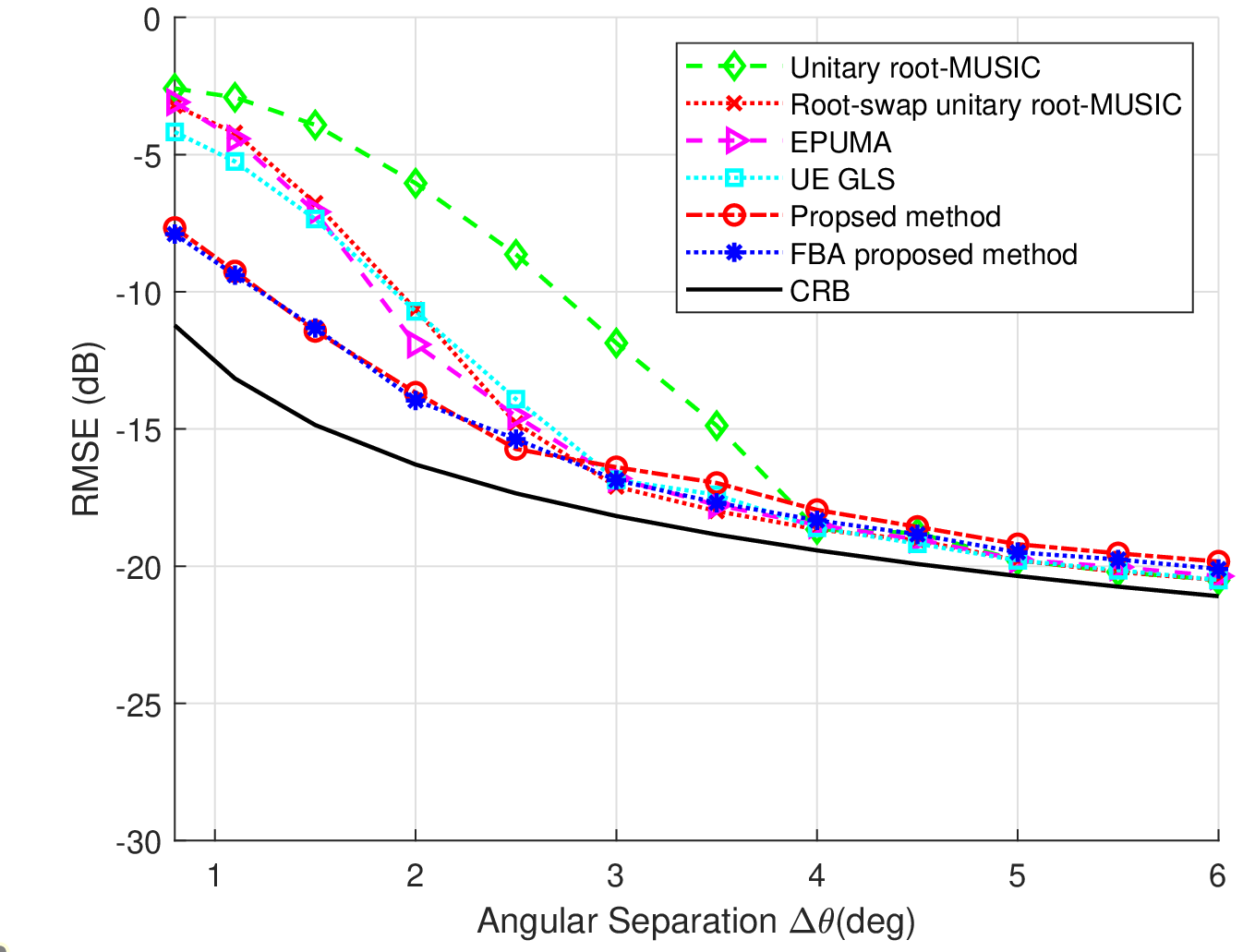}}
		\vspace{-0.5cm}
		\caption{ RMSE vs. the angular separation for $L=3$ uncorrelated sources with $\boldsymbol{\theta} = [0^{\circ} , 34^{\circ},  (34+\Delta\theta)^{\circ}]$, SNR~$ =10$~dB, $M=10$, and $N=10$.}
		\label{fig6}
	\end{center}
\end{figure}

\begin{figure}[!]
	\begin{center}
		{\includegraphics[width=3.5in]{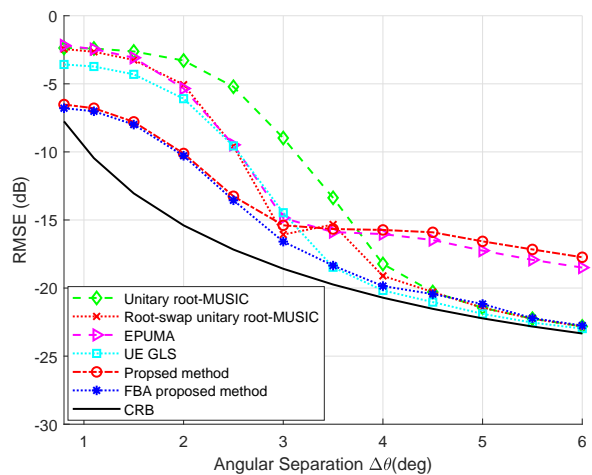}}
		\vspace{-0.5cm}
		\caption{ RMSE vs. the angular separation for $L=3$ partly correlated sources with $\boldsymbol{\theta} = [0^{\circ}, 34^{\circ},  (34+\Delta\theta)^{\circ}]$, SNR~$ =15$~dB, $\rho=0.95$ for the last two directions, $M=10$, and $N=10$.}
		\label{fig7}
	\end{center}
\end{figure}

In the first example, three uncorrelated sources located at $\boldsymbol{\theta} = [19^{\circ},  34^{\circ}, 36^{\circ}]$ are considered. It can be seen in Fig.~\ref{fig2} that the SNR threshold performance of the proposed method is outstandingly better than that of the other methods tested. In Fig.~\ref{fig3}, the capability of different methods to cope with correlated sources is evaluated by letting $\boldsymbol{\theta} = [5^{\circ}  ,  8^{\circ}]$ be correlated with the correlation coefficient $\rho = 0.95$. As expected, the performance of the forward-only version of the proposed method degrades in relatively high SNR region, while it is as good as that of the FBA version of the proposed method in low SNRs. In addition, the FBA version of the proposed method possesses much better performance compared to other methods. As illustrated in Fig.~\ref{fig4}, the impact of the number of snapshots on the performance of the methods tested is investigated via setting $\boldsymbol{\theta} = [10^{\circ},  34^{\circ}, 36^{\circ}]$, $\rho=0$ for the fixed SNR~$=5$~dB. Similarly, Fig.~\ref{fig5} shows the impact of the number of snapshots for a different setup of $\boldsymbol{\theta} = [34^{\circ},  38^{\circ}]$, $\rho=0.95$, and SNR~$=2$~dB. It displays that the FBA version of the proposed method provides robust estimates even when the number of snapshots is about one order of magnitude smaller than that of the other methods tested. In the next setup, the capability of the methods tested to deal with the scenario of two closely located sources is investigated. In doing so, we regard the setup in which $\boldsymbol{\theta} = [0^{\circ}, 34^{\circ},  (34+\Delta\theta)^{\circ}]$, $\rho=0$, and SNR~$=10$~dB with $\Delta\theta$ varying from $0.8^{\circ}$ to $6^{\circ}$. It can be seen in Fig.~\ref{fig6} that the proposed method is more reliable in dealing with smaller angular separations compared to the other methods tested. The results for the same setup as Fig.~\ref{fig6}, but with $\rho=0.95$ for the last two directions, and SNR~$=15$~dB, are shown Fig.\ref{fig7}. It can be seen that the FBA version of the proposed method has the best performance.


\begin{figure}[!]
	\begin{center}
		{\includegraphics[width=3.5in]{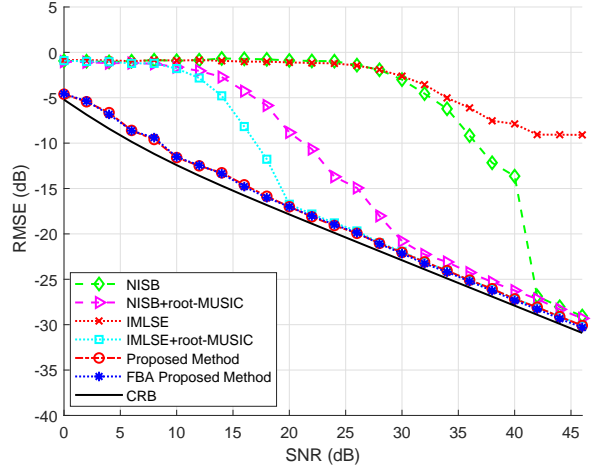}}
		\vspace{-0.5cm}
		\caption{ RMSE vs. SNR for $L=2$ uncorrelated sources with $\boldsymbol{\theta} = [ 33^{\circ}, 36^{\circ}]$, $M=8$, and $N=10$.}
		\label{fig8}
	\end{center}
\end{figure}

\begin{figure}[!]
	\begin{center}
		{\includegraphics[width=3.5in]{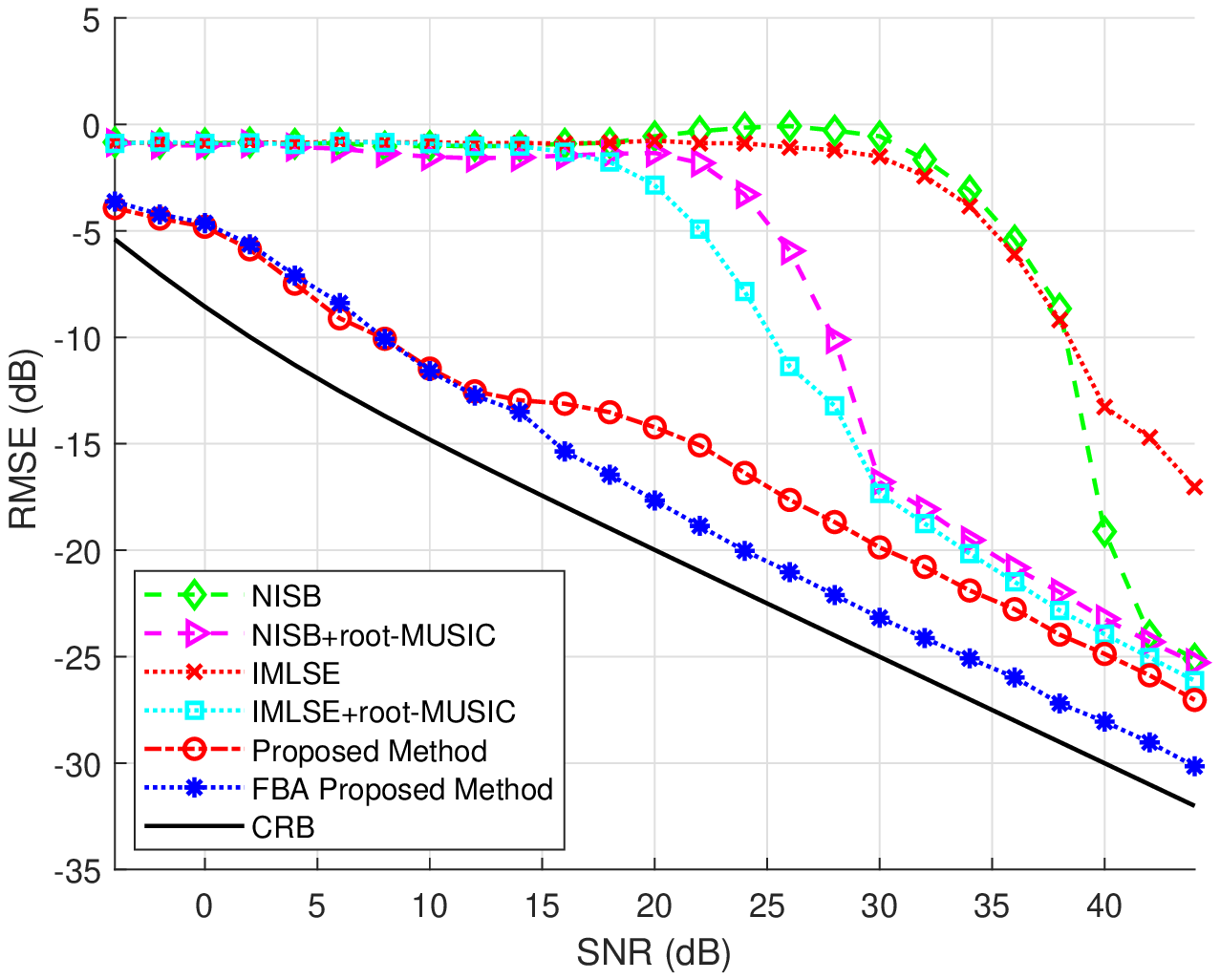}}
		\vspace{-0.5cm}
		\caption{ RMSE vs. SNR for $L=2$ correlated sources with $\boldsymbol{\theta} = [ 33^{\circ}, 38^{\circ}]$, $\rho=0.95$, $M=8$, and $N=10$.}
		\label{fig9}
	\end{center}
\end{figure}

\begin{figure}[!]
	\begin{center}
		{\includegraphics[width=3.5in]{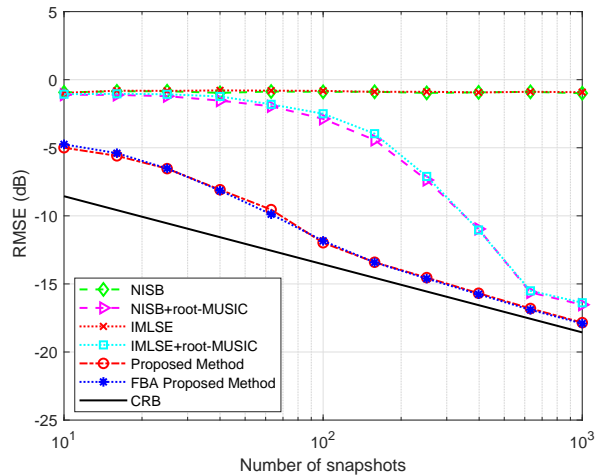}}
		\vspace{-0.5cm}
		\caption{ RMSE vs. the number of snapshots for $L=2$ uncorrelated sources with $\boldsymbol{\theta} = [ 33^{\circ}, 38^{\circ}]$, SNR~$ =0$~dB, and $M=8$.}
		\label{fig10}
	\end{center}
\end{figure}

\begin{figure}[!]
	\begin{center}
		{\includegraphics[width=3.5in]{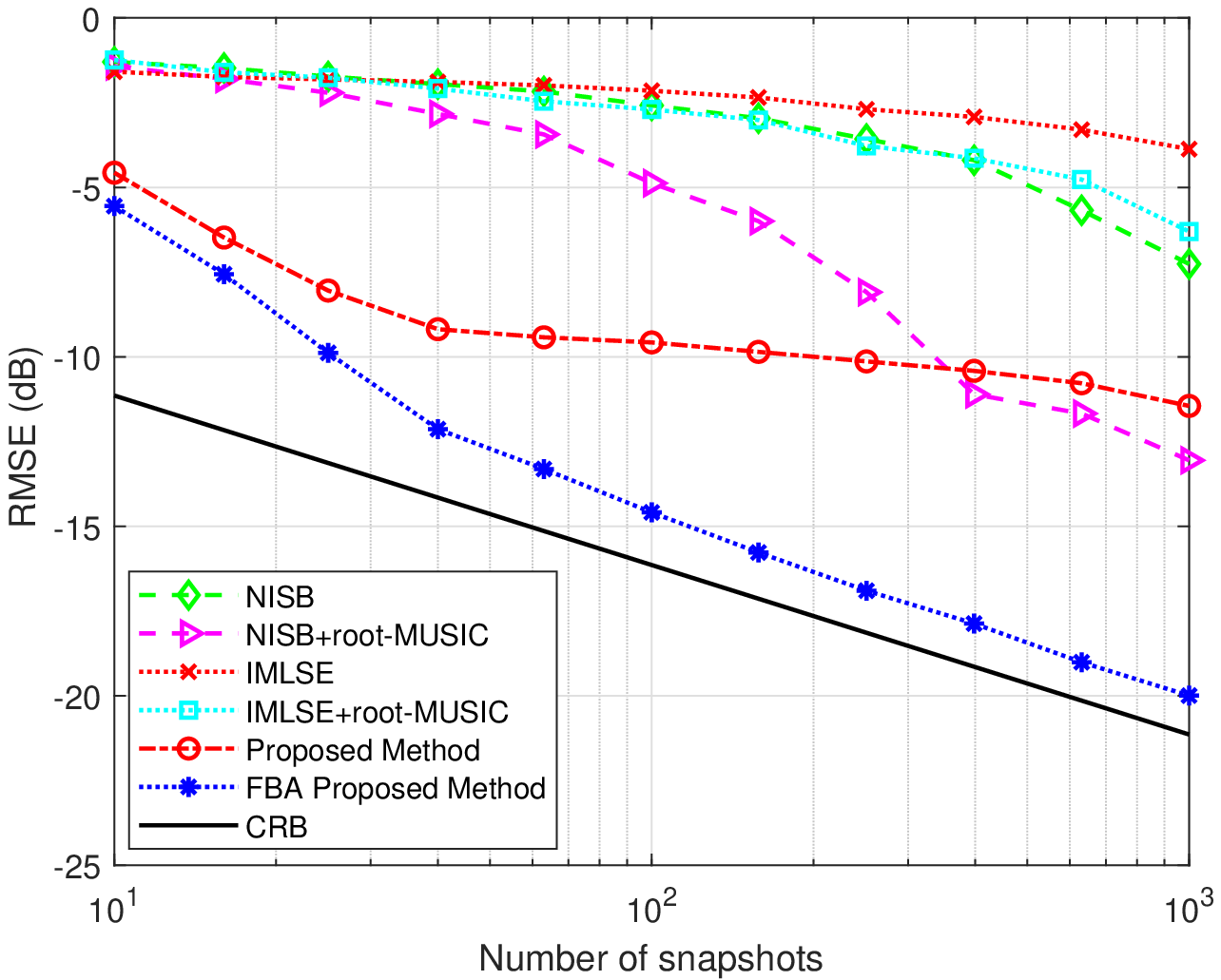}}
		\vspace{-0.5cm}
		\caption{ RMSE vs. the number of snapshots for $L=2$ correlated sources with $\boldsymbol{\theta} = [ 33^{\circ}, 48^{\circ}]$, $\rho=0.95$, SNR~$ =-4$~dB, and $M=8$.}
		\label{fig11}
	\end{center}
\end{figure}

\begin{figure}[!]
	\begin{center}
		{\includegraphics[width=3.5in]{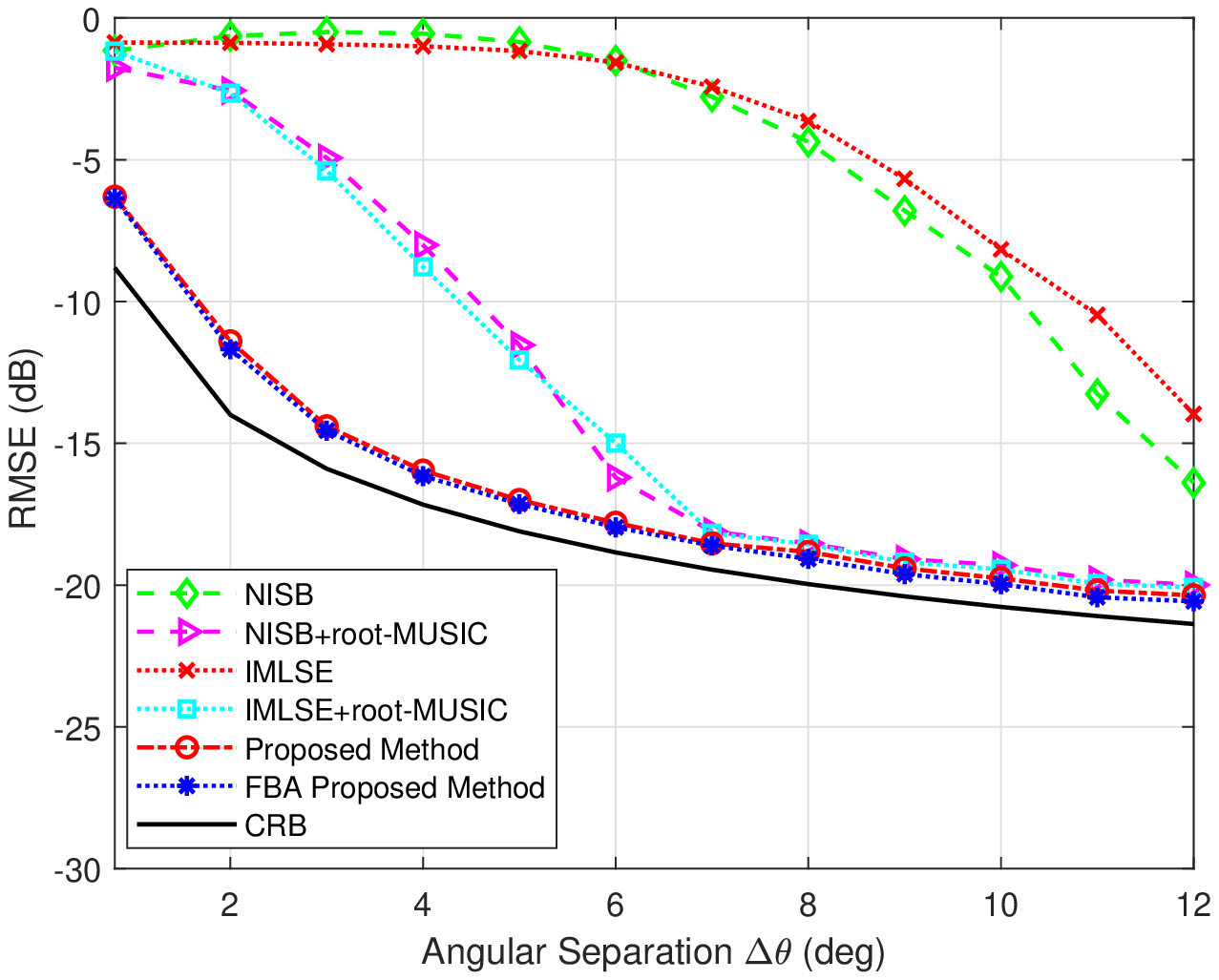}}
		\vspace{-0.5cm}
		\caption{ RMSE vs. the Angular Separation for $L=3$ uncorrelated sources with $\boldsymbol{\theta} = [-10^{\circ} , 34^{\circ}  ,  (34+\Delta\theta)^{\circ}]$, SNR~$ =15$~dB, $M=8$, and $N=10$.}
		\label{fig12}
	\end{center}
\end{figure}

\begin{figure}[!]
	\begin{center}
		{\includegraphics[width=3.5in]{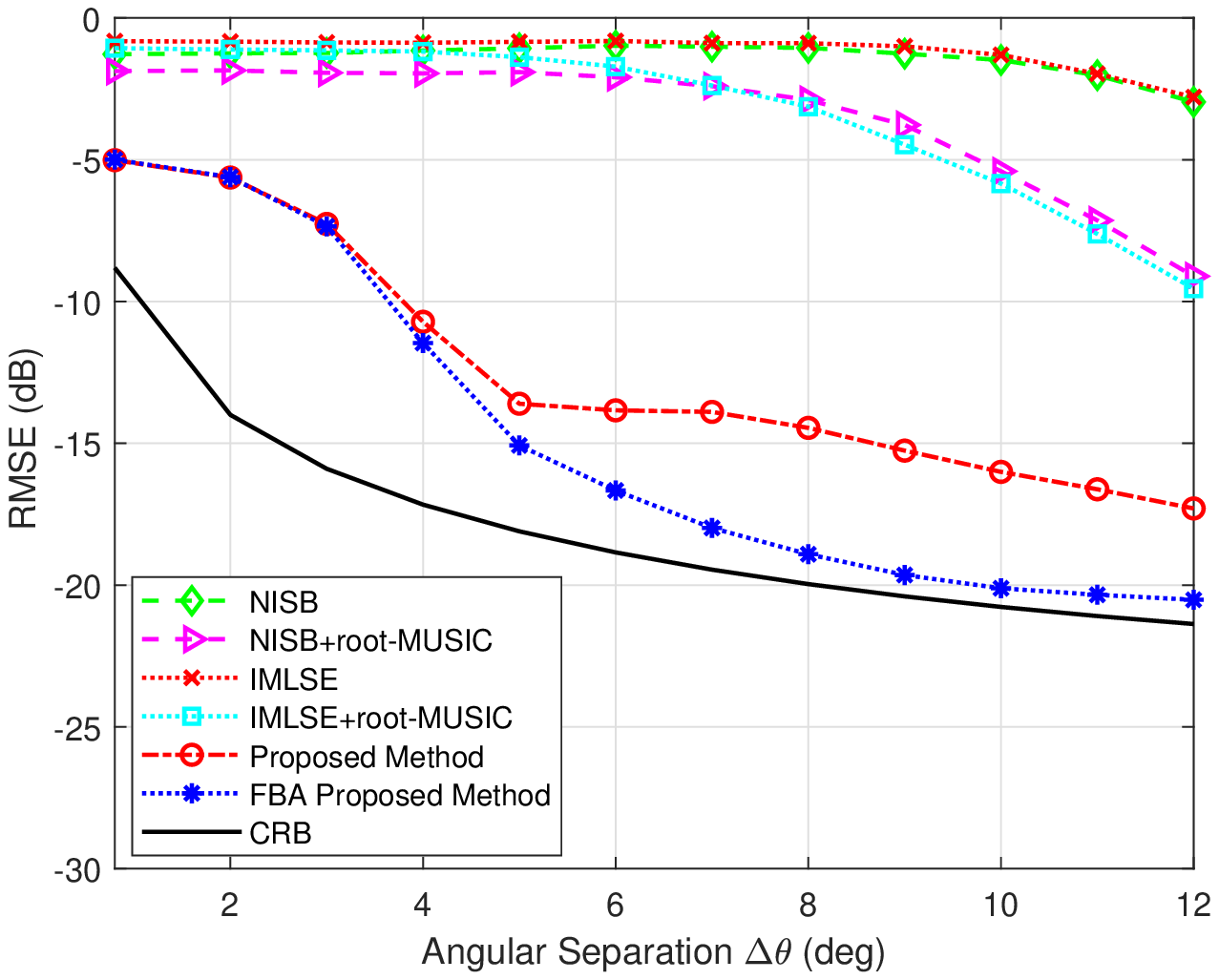}}
		\vspace{-0.5cm}
		\caption{ RMSE vs. the Angular Separation for $L=3$ partly correlated sources with $\boldsymbol{\theta} = [-10^{\circ} , 34^{\circ},  (34+\Delta\theta)^{\circ}]$, SNR~$ =15$~dB, $\rho=0.95$ for the last two directions, $M=8$, and $N=10$.}
		\label{fig13}
	\end{center}
\end{figure}

To study the impact of the nonuniform noise, the sensor noise covariance matrix is set as $\mathbf{Q}=\mathrm{diag}\{[6, 2, 0.5, 2.5, 3, 1, 5.5, 10]\}$ for the follow up examples \cite{liao2016iterative}. The worst noise power ratio (WNPR) used in these examples is given as
\begin{flalign}
	\mathrm{WNPR}= \frac{\sigma_{max}^2}{\sigma_{min}^2} = \frac{10}{0.5} = 20 . \nonumber
\end{flalign}
First, Fig.~\ref{fig8} shows the RMSE performance of the methods tested versus SNR for the setup of $\boldsymbol{\theta} = [33^{\circ}, 36^{\circ}]$ and $\rho=0$. As it can be observed, the threshold performance of the proposed methods is substantially better than that of the other methods tested. Moreover, in Fig.~\ref{fig9}, the same scenario as that shown in Fig.~\ref{fig8} is considered for the case of correlated sources $\boldsymbol{\theta} = [ 33^{\circ}, 38^{\circ}]$ and $\rho=0.95$. Fig.~\ref{fig9} demonstrates the superiority of the FBA version of the proposed method over the other methods tested. Fig.~\ref{fig10} depicts how  different methods perform depending on the number of snapshots for the scenario of $\boldsymbol{\theta} = [ 33^{\circ}, 38^{\circ}]$, $\rho=0$, and SNR~$=0$~dB. Fig.~\ref{fig11} illustrates this dependency also for the scenario of $\boldsymbol{\theta} = [ 33^{\circ}, 48^{\circ}]$, $\rho=0.95$, and SNR~$=-4$~dB. Based on Figs.~\ref{fig10}~and~\ref{fig11}, it can be concluded that the reliability of the proposed method to the scarcity of the number of snapshots is higher for the case of uncorrelated sources compared to the other methods tested, while the FBA version of the proposed method copes with the correlated sources more efficiently. Finally, Figs.~\ref{fig12}~and~\ref{fig13} show the strengths of the methods tested against the presence of closely located sources for the cases of uncorrelated and correlated signals, respectively. The setup regarded for Fig.~\ref{fig12} is $\boldsymbol{\theta} = [-10^{\circ} , 34^{\circ},  (34+\Delta\theta)^{\circ}]$, $\rho=0$, and SNR~$=15$~dB with $\Delta\theta$ varying from $1^{\circ}$ to $12^{\circ}$. It can be observed that the performance of the proposed method almost achieves the CRB. Fig.~\ref{fig13} depicts the results obtained from conducting the same setup as for Fig.~\ref{fig12} with the difference that $\rho=0.95$ for the last two directions. The superiority of the FBA version of the proposed method over other methods tested can be seen to be very substantial.

\section{Conclusion}
\label{sec6}
A new algorithm for DOA estimation in the presence of nonuniform sensor noise is introduced. The proposed algorithm works in three phases. The essence of the first phase is to estimate the nonuniform noise covariance matrix in an iterative manner. In each iteration, the noise subspace is estimated using GED first and then the noise covariance estimate is updated as the solution of an LS minimization problem. The asymptotic performance of one iteration of such algorithm is investigated. The advantage of the proposed noise covariance estimator is that it is applicable to any array geometry. After the noise covariance matrix is estimated, candidate DOAs are estimated using a rooting-based DOA estimation method based on the combination of the GLS and the first-order signal subspace perturbation. The noise covariance estimate obtained in the first phase is used for pre-whitening the array signal with nonuniform noise. In addition, the forward-only DOA estimation method is extended using  FBA. Furthermore, the asymptotic performance of both the forward-only and FBA versions of the proposed method is studied. In the third phase, the final best $L$ DOA estimates out of $2L$ DOA candidates generated in the second phase are selected using properly designed DOA selection strategy. Numerical simulation examples are included to show the superiority of the proposed algorithm compared to the state-of-the-art approaches for both cases of the uniform and nonuniform sensor noise.      

%

\appendices
\section{}
\label{APP1}
Taking into account the fact that $\mathbf{Q}$ is a real-valued diagonal matrix and also using the derivative properties \cite{petersen2012matrix}, the partial derivatives of the terms related to $\mathbf{Q}$ in \eqref{eq310} can be found to be  
\begin{flalign}
	\frac{\partial \mathrm{trace} \left\{ \hat{\mathbf{R}} \hat{\mathbf{U}} \hat{\mathbf{U}}^H \mathbf{Q} \right\}}{\partial \mathbf{Q}} &= \frac{\partial \mathrm{trace} \left\{ \mathcal{D} \left\{ \hat{\mathbf{R}} \hat{\mathbf{U}} \hat{\mathbf{U}}^H \right\} \mathbf{Q} \right\}}{\partial \mathbf{Q}} 
	= \mathcal{D} \left\{ \hat{\mathbf{R}} \hat{\mathbf{U}} \hat{\mathbf{U}}^H \right\}
	\label{eqA1} \\
	\frac{\partial \mathrm{trace} \left\{  \hat{\mathbf{U}} \hat{\mathbf{U}}^H  \hat{\mathbf{R}}  \mathbf{Q} \right\}}{\partial \mathbf{Q}} &= \frac{\partial \mathrm{trace} \left\{ \mathcal{D} \left\{  \hat{\mathbf{U}} \hat{\mathbf{U}}^H   \hat{\mathbf{R}}    \right\} \mathbf{Q} \right\}}{\partial \mathbf{Q}} 
	= \mathcal{D} \left\{  \hat{\mathbf{U}} \hat{\mathbf{U}}^H  \hat{\mathbf{R}}  \right\}
	\label{eqA2} \\
	\frac{\partial \mathrm{trace} \left\{  \hat{\mathbf{U}} \hat{\mathbf{U}}^H   \mathbf{Q}^2 \right\}}{\partial \mathbf{Q}} &= \frac{\partial \mathrm{trace} \left\{ \mathcal{D} \left\{  \hat{\mathbf{U}} \hat{\mathbf{U}}^H   \right\} \mathbf{Q}^2 \right\}}{\partial \mathbf{Q}} 
	= 2 \mathcal{D} \left\{  \hat{\mathbf{U}} \hat{\mathbf{U}}^H  \right\} \mathbf{Q}.
	\label{eqA3}
\end{flalign}
Using \eqref{eqA1}--\eqref{eqA3}, the partial derivative of \eqref{eq310} with respect to $\mathbf{Q}$ can be straightforwardly found to be \eqref{eq311}.

\section{}
\label{APP2}
Using \eqref{eq313}, the $m$th diagonal entry of $\hat{\mathbf{Q}}$, denoted by $\hat{\sigma}_m^2$, can be written as
\begin{flalign}
	\hat{\sigma}_m^2 = \frac{\left(\mathbf{v}_m^H \hat{\mathbf{r}}_m + (\mathbf{v}_m^H \hat{\mathbf{r}}_m )^H \right)}{2 \tau_m}  = \frac{ \mathfrak{R} \{\mathbf{v}_m^H \hat{\mathbf{r}}_m \} }{ \tau_m}  
	\label{eqB1} 
\end{flalign}
where $\tau_m \triangleq [\hat{\mathbf{U}} \hat{\mathbf{U}}^H]_{mm}$, and $\mathbf{v}_m \triangleq [\hat{\mathbf{U}} \hat{\mathbf{U}}^H]_{:,m}$. Expressing $\hat{\mathbf{r}}_m$ as $\hat{\mathbf{r}}_m = \mathbf{r}_m + \boldsymbol{\Delta} \mathbf{r}_m$, where $\boldsymbol{\Delta} \mathbf{r}_m$ denotes the estimation error of the $m$th column of the SCM, it can be written that
\begin{flalign}
	\Delta\sigma_m^2 =  \frac{ \mathfrak{R} \{\mathbf{v}_m^H  \boldsymbol{\Delta} \mathbf{r}_m \} }{ \tau_m}  
	\label{eqB2} 
\end{flalign}
where $\Delta\sigma_m^2$ is the difference between the actual $\sigma_m^2$ and the estimate $\hat{\sigma}_m^2$, i.e., $\Delta\sigma_m^2 = \hat{\sigma}_m^2 - \sigma_m^2$. As a result, the variance of $\Delta\sigma_m^2$ can be expressed as
\begin{flalign}
	\mathbb{E} \left\{ \left(\Delta \sigma_m^2 \right)^2 \right\} &= \frac{1}{4 \tau_m^2} \mathbb{E} \bigg\{ \left( \mathbf{v}_m^H \boldsymbol{\Delta} \mathbf{r}_m + \mathbf{v}_m^T \boldsymbol{\Delta} \mathbf{r}_m^* \right)  \left(  \boldsymbol{\Delta} \mathbf{r}_m^H \mathbf{v}_m + \boldsymbol{\Delta} \mathbf{r}_m^T \mathbf{v}_m^* \right) \bigg\} \nonumber \\ 
	&= \frac{1}{4 \tau_m^2} \bigg( \mathbf{v}_m^H \mathbb{E} \left\{ \boldsymbol{\Delta} \mathbf{r}_m \boldsymbol{\Delta} \mathbf{r}_m^H \right\} \mathbf{v}_m + \mathbf{v}_m^H \mathbb{E} \left\{ \boldsymbol{\Delta} \mathbf{r}_m \boldsymbol{\Delta} \mathbf{r}_m^T \right\} \mathbf{v}_m^* \nonumber \\
	&+ \mathbf{v}_m^T \mathbb{E} \left\{ \boldsymbol{\Delta} \mathbf{r}_m^* \boldsymbol{\Delta} \mathbf{r}_m^H \right\} \mathbf{v}_m + \mathbf{v}_m^T \mathbb{E} \left\{ \boldsymbol{\Delta} \mathbf{r}_m^* \boldsymbol{\Delta} \mathbf{r}_m^T \right\} \mathbf{v}_m^* \bigg). 
	\label{eqB4}
\end{flalign}
According to \cite{bilodeau2008theory}, the asymptotic covariance and pseudo-covariance matrices of the vector $\boldsymbol{\Delta} \mathbf{r} \triangleq \mathrm{vec} \{ (\hat{\mathbf{R}} - \mathbf{R})\} \in \mathbb{C}^{M^2 }$ are
\begin{flalign}
	\mathbb{E} \left\{ \boldsymbol{\Delta} \mathbf{r} \boldsymbol{\Delta} \mathbf{r}^H \right\} &= \frac{1}{N} (\mathbf{R}^T \otimes \mathbf{R}) 
	\label{eqB5} \\
	\mathbb{E} \left\{ \boldsymbol{\Delta} \mathbf{r} \boldsymbol{\Delta} \mathbf{r}^T \right\} &= \mathbf{R}^T \otimes \mathbf{R}. 
	\label{eqB6}
\end{flalign}
Using \eqref{eqB5} and \eqref{eqB6}, it is straightforward to show that \cite{ottersten1998covariance}
\begin{flalign}
	\mathbb{E} \left\{ \boldsymbol{\Delta} \mathbf{r}_m \boldsymbol{\Delta} \mathbf{r}_m^H \right\} &= \left( \frac{[\mathbf{R}]_{mm}}{N} \right) \mathbf{R} 
	\label{eqB7} \\
	\mathbb{E} \left\{ \boldsymbol{\Delta} \mathbf{r}_m \boldsymbol{\Delta} \mathbf{r}_m^T \right\} &= ([\mathbf{R}]_{mm}) \mathbf{R}.
	\label{eqB8}
\end{flalign}

Plugging \eqref{eqB7} and \eqref{eqB8} into \eqref{eqB4} yields
\begin{flalign}
	\mathbb{E} \left\{ \left(\Delta \sigma_m^2 \right)^2 \right\} &= \left( \frac{ [\mathbf{R}]_{mm}}{2 N \tau_m^2} \right) \mathfrak{R} \left\{ \mathbf{v}_m^H \mathbf{R} \left( \mathbf{v}_m + N \mathbf{v}_m^* \right)\right\}
	\label{eqB9} 
\end{flalign}
which completes the proof.

\section{}
\label{APP3}
The proof goes in the same steps as that in \cite{qian2016enhanced} (Appendix~A), and is included for the sake of completeness.
As $\theta_l$ and $\gamma_l$ are related to each other as $\gamma_l = e^{-j 2 \pi d \mathrm{sin}(\theta_l) / \lambda}$, we perform Taylor's expansion and keep only the terms containing up to the first-order perturbation terms to obtain    
\begin{flalign}
	\Delta \theta_l \approx - \frac{\lambda}{2 \pi d \cos(\theta_l)} \frac{\Delta \gamma_l}{j \gamma_l} . 
	\label{eqC1} 
\end{flalign}
To enforce $\Delta \theta_l$ to be a real-valued quantity, it is reasonable to define
\begin{flalign}
	\Delta \theta_l &\triangleq  \frac{1}{2} (\Delta \theta_l + \Delta \theta_l^*) = \frac{1}{2} \frac{j \lambda}{2 \pi d \cos(\theta_l)} (\gamma_l^* \Delta\gamma_l - \gamma_l \Delta\gamma_l^*).
	\label{eqC2}
\end{flalign}
Using \eqref{eqC2}, the variance of $\Delta \theta_l$ can be written as
\begin{flalign}
	\mathbb{E} \{ \Delta \theta_l^2 \} &\approx  \frac{1}{2} \left(\frac{ \lambda}{2 \pi d \cos(\theta_l)}\right)^2  \left( \mathbb{E} \{ | \Delta \gamma_l |^2 \} - \mathfrak{R} \{ \mathbb{E} \{  \Delta \gamma_l^2 \} (\gamma_l^*)^2\} \right).
	\label{eqC3}
\end{flalign}

Next we need to find an expression that connects $\Delta \gamma_l$ and $\boldsymbol{\Delta} \mathbf{a} \triangleq \hat{\mathbf{a}}-\mathbf{a}$, as \eqref{eqC3} is dependent to the statistics of $\Delta \gamma_l$ which are related to the statistics of $\boldsymbol{\Delta} \mathbf{a}$. Towards this end, using the first-order approximation of \eqref{eq343} when $\gamma$ is replaced by $\gamma_l$, we obtain 
\begin{flalign}
	\boldsymbol{\gamma}_l^T \boldsymbol{\Delta}\mathbf{a} + \phi_l \Delta \gamma_l \approx 0
	\label{eqC4}
\end{flalign}
where $\boldsymbol{\Delta}\mathbf{a} \triangleq [\Delta [\mathbf{a}]_1 \cdots \Delta [\mathbf{a}]_L]^T$, $\boldsymbol{\gamma}_l \triangleq [\gamma_l^{L-1} \cdots 1]^T$, $\phi_l \triangleq L \gamma_l^{L-1} + (L-1) [\mathbf{a}]_1 \gamma_l^{L-2} + \cdots + [\mathbf{a}]_{L-1}$.
From \eqref{eqC4}, we obtain $\Delta \gamma_l \approx -\frac{\boldsymbol{\gamma}_l^T \mathbf{\Delta} \mathbf{a}}{\phi_l}$. Thus, it can be written that 
\begin{flalign}
	\mathbb{E} \{ |\Delta \gamma_l |^2 \} \approx \frac{\boldsymbol{\gamma}_l^T \mathbb{E} \{ \boldsymbol{\Delta}\mathbf{a} \boldsymbol{\Delta}\mathbf{a}^H  \} \boldsymbol{\gamma}_l^*}{| \phi_l|^2}.
	\label{eqC8}
\end{flalign}

Let us define
\begin{flalign}
	f(\mathbf{a}) = (\hat{\mathbf{H}} \mathbf{a} - \hat{\mathbf{h}} )^H \mathbf{W} (\hat{\mathbf{H}} \mathbf{a} - \hat{\mathbf{h}} ).
	\label{eqC9}
\end{flalign}
Since $\hat{\mathbf{a}}$ is the vector that minimizes \eqref{eqC9}, $f'(\hat{\mathbf{a}})$ can be approximated under the assumption of high SNR using Taylor's expansion as \cite{qian2016enhanced}, \cite{so2013simple}  
\begin{flalign}
	0 = f'(\hat{\mathbf{a}}) \approx f'(\mathbf{a}) + f''(\mathbf{a}) \boldsymbol{\Delta} \mathbf{a}
	\label{eqC10}
\end{flalign}
where $f'(\mathbf{a})$ and $f''(\mathbf{a})$ denote respectively the first and second derivatives of $f(\mathbf{a})$ with respect to $\mathbf{a}$, which are given as 
\begin{flalign}
	f'(\mathbf{a}) &= 2 \hat{\mathbf{H}}^H \mathbf{W} (\hat{\mathbf{H}} \mathbf{a} - \hat{\mathbf{h}} ) = 2 \hat{\mathbf{H}}^H \mathbf{W} \hat{\mathbf{e}}= 2 \hat{\mathbf{H}}^H \mathbf{W} \left(\mathbf{I}_L \otimes \mathbf{C}(\mathbf{a})\right) \boldsymbol{\Delta} \mathbf{u}_s 
	\label{eqC11} \\
	f''(\mathbf{a}) &= 2 \hat{\mathbf{H}}^H \mathbf{W} \hat{\mathbf{H}}.
	\label{eqC12}
\end{flalign}

Combining \eqref{eqC10}, \eqref{eqC11} and \eqref{eqC12}, we get for high SNR that
\begin{flalign}
	\mathbb{E} &\{ \boldsymbol{\Delta}\mathbf{a}  \boldsymbol{\Delta}\mathbf{a}^H \} \approx  (\mathbf{H}^H \mathbf{W} \mathbf{H})^{-1} \mathbf{H}^H \mathbf{W} \left(\mathbf{I}_L \otimes \mathbf{C}(\mathbf{a})\right) \nonumber \\ 
	&\times \mathbb{E} \{ \boldsymbol{\Delta} \mathbf{u}_{\rm s} \boldsymbol{\Delta}  \mathbf{u}_{\rm s}^H\} 
	\left(\mathbf{I}_L \otimes \mathbf{C}^H (\mathbf{a}) \right) \mathbf{W} \mathbf{H} (\mathbf{H}^H \mathbf{W} \mathbf{H})^{-1}.
	\label{eqC13} 
\end{flalign}
Based on \eqref{eq370} and \eqref{eqC13}, it can be written that 
\begin{flalign}
	\mathbb{E} \{ \boldsymbol{\Delta}\mathbf{a}  \boldsymbol{\Delta}\mathbf{a}^H \} \approx  (\hat{\mathbf{H}}^H \mathbf{W} \hat{\mathbf{H}})^{-1} .
	\label{eqC14} 
\end{flalign}
Consequently, substituting \eqref{eqC14} into \eqref{eqC8}, we have
\begin{flalign}
	\mathbb{E} \{ |\Delta \gamma_l |^2 \} \approx \frac{\boldsymbol{\gamma}_l^T (\hat{\mathbf{H}}^H \mathbf{W} \hat{\mathbf{H}})^{-1} \boldsymbol{\gamma}_l^*}{| \phi_l|^2}.
	\label{eqC15} 
\end{flalign}

The final part is to compute $ \mathbb{E} \{  \Delta \gamma_l^2 \}$, which has the following form
\begin{flalign}
	\mathbb{E} \{ \Delta \gamma_l^2 \} \approx \frac{\boldsymbol{\gamma}_l^T \mathbb{E} \{ \boldsymbol{\Delta}\mathbf{a} \boldsymbol{\Delta}\mathbf{a}^T  \} \boldsymbol{\gamma}_l}{ \phi_l^2}
	\label{eqC16}
\end{flalign}
where
\begin{flalign}
\mathbb{E} \{ \boldsymbol{\Delta}\mathbf{a}  \boldsymbol{\Delta}\mathbf{a}^T \} &\approx  (\mathbf{H}^H \mathbf{W} \mathbf{H})^{-1} \mathbf{H}^H \mathbf{W} \left(\mathbf{I}_L \otimes \mathbf{C}(\mathbf{a})\right) \nonumber \\  &\times \mathbb{E} \{ \boldsymbol{\Delta} \mathbf{u}_{\rm s} \boldsymbol{\Delta}  \mathbf{u}_{\rm s}^T\} 
\left(\mathbf{I}_L \otimes \mathbf{C}(\mathbf{a})^T \right) \mathbf{W}^T \mathbf{H}^* (\mathbf{H}^H \mathbf{W} \mathbf{H})^{-T}.
\label{eqC17}
\end{flalign}
It follows from \eqref{eq372} that  
\begin{flalign}
	\mathbb{E} \{ \boldsymbol{\Delta} \mathbf{u}_{\rm s} \boldsymbol{\Delta}  \mathbf{u}_{\rm s}^T\} \approx \left( \boldsymbol{\Sigma}_{\rm s}^{-1} \mathbf{V}_{\rm s}^T \otimes (\mathbf{I}_M - \mathbf{U}_{\rm s} \mathbf{U}_{\rm s}^H) \right) \mathbb{E} \{ \Bar{\mathbf{n}} \Bar{\mathbf{n}}^T \} \times \left(  \mathbf{V}_{\rm s} \boldsymbol{\Sigma}_{\rm s}^{-1} \otimes (\mathbf{I}_M - \mathbf{U}_{\rm s}^* \mathbf{U}_{\rm s}^T) \right).
	\label{eqC18} 
\end{flalign}
Since $\mathbb{E} \{ \Bar{\mathbf{n}} \Bar{\mathbf{n}}^T \} = \mathbf{0}_{MN \times MN}$, \eqref{eqC18} becomes a zero matrix which gives rise to
\begin{flalign}
	\mathbb{E} \{ \Delta \gamma_l^2 \} \approx 0 .
	\label{eqC19} 
\end{flalign}
As a result, by combining \eqref{eqC3}, \eqref{eqC15} and \eqref{eqC19}, we obtain \eqref{eq396}, which completes the proof.

\ifCLASSOPTIONcaptionsoff
  \newpage
\fi



%
\bibliographystyle{IEEEtran}
\bibliography{IEEEabrv,refs}


%








\end{document}